%% file: main.tex
\documentclass[1p]{elsarticle}

\usepackage{graphicx,url}
\usepackage[figtopcap]{subfigure}
\usepackage[utf8]{inputenc}  
\usepackage{float}
\usepackage{lineno}
\usepackage{pslatex}
\usepackage[scaled=0.92]{helvet}
\usepackage[pdfpagelabels]{hyperref}
\usepackage{amsmath,amsthm,amsfonts,amssymb}
\usepackage[mathcal]{eucal}
\usepackage{latexsym}
\usepackage{textcase}
\usepackage[printonlyused]{acronym}
\usepackage{bm}
\usepackage{multirow}
\usepackage{booktabs}
\usepackage{tabulary}
\usepackage{longtable}
\usepackage{subfigure}
\usepackage{threeparttable}

\journal{Computers \& Security}
\begin{document} 
\newcommand{\citeonline}{\cite}
\newcommand{\fontDFD}[1]{\textsf{#1}}
\newcommand{\fontArquivos}[1]{\textsf{#1}}
\newcommand{\Tag}[1]{\emph{#1}}  

\begin{frontmatter}

\title{An Integrity--Focused Threat Model for Software Development Pipelines}

\author[ppgcap]{Beatriz Michelson Reichert and Rafael R.\ Obelheiro}
\address{Graduate Program in Applied Computing, State University of Santa Catarina}
\begin{abstract}
\input{chapters/0_Abstract}
\end{abstract}

\end{frontmatter}

\input{chapters/1_Introduction}
\input{chapters/2_Background}

\input{chapters/3_RelatedWork}

\input{chapters/4_ThreatModeling}
\input{chapters/5_CaseStudy}
\input{chapters/6_Conclusion}

\section*{Acknowledgments}

Funding: this work was supported by FAPESC and UDESC\@. 


\bibliographystyle{elsarticle-harv}
\bibliography{tcc-biblio}

\end{document}

%% file: chapters/0_Abstract.tex
In recent years, there has been a growing concern with software integrity, that is, the assurance that software has not been tampered with on the path between developers and users. This path is represented by a software development pipeline and plays a pivotal role in software supply chain security. While there have been efforts to improve the security of development pipelines, there is a lack of a comprehensive view of the threats affecting them. We develop a systematic threat model for a generic software development pipeline using the STRIDE framework and identify possible mitigations for each threat. The pipeline adopted as a reference comprises five stages (integration, continuous integration, infrastructure-as-code, deployment, and release), and we review vulnerabilities and attacks in all stages reported in the literature. We present a case study applying this threat model to a specific pipeline, showing that the adaptation is straightforward and produces a list of relevant threats.

%% file: chapters/1_Introduction.tex
\section{Introduction} \label{cap:introdução}

For the past quarter century, software security has been on the agenda for both industry and academia \cite{mcgraw2012software}. The ever-growing reliance of society on software makes software security a priority. Over this time, developing secure software has been the foremost concern, with a focus on identifying security requirements, writing bug- and vulnerability-free code, and providing mechanisms to  configure systems securely \cite{allen2008software,mcgraw2012software,hamid2018engineering}.

A recent trend is a heightened concern with the security of the software supply chain \cite{enck2022top,enisa2021threat}. Noteworthy incidents such as the ones involving malicious Avast CCleaner \cite{brumaghin2017} and SolarWinds Orion \cite{goodin2020solarwinds,peisert2021perspectives} software updates have raised awareness about the risk of vulnerabilities and malicious logic being surreptitiously introduced along the path between developers and users. A software supply chain may refer to the process and infrastructure used by an organization to build and deliver a single software product \cite{torres2019toto}, or to multi-tiered relationships among organizations, where an organization produces some component that is used by another organization in its development process (for instance, when organization~\emph{A} develops an application that uses libraries developed by vendors~\emph{B} and~\emph{C}) \cite{simpson2010software}. The infrastructure used by a single organization is also called a software development pipeline \cite{adams2016modern}. To avoid ambiguity, in this paper we use ``development pipeline'' to refer to a single organization, and ``supply chain'' to refer to multi-tiered relationships among organizations. Thus, a software supply chain can be viewed as interconnected development pipelines, where the output of one pipeline may be delivered to end users or serve as input to another pipeline.

Ensuring users receive software products as intended by their developers, without vulnerabilities being introduced along the way, demands secure software development pipelines. There have been some initiatives on software supply chain/development pipeline security from both industry and academia. On the industry side, \citeonline{simpson2010software} discusses several integrity practices and controls for software supply chains. Supply chain threats and countermeasures in the context of information technology acquisitions (particularly in the defense sector) have been cataloged in \cite{miller2013supply,he2015model}. Publicized attacks on development pipelines between 2014 and 2017 are summarized in \cite{shaw}. A report from ENISA \cite{enisa2021threat} introduces a taxonomy for supply chain attacks and reviews 24~attacks between January 2020 and July 2021. Supply chain security for open source software is addressed in \cite{thelinuxfoundation2020}. The security of build and deployment infrastructures is part of software security/assurance maturity models such as the Building Security In Maturity Model (BSIMM) \cite{BSIMM2022} 
and the OWASP Security Assurance Maturity Model (SAMM) \cite{owaspSAMM2020}, as well as 
the ISO/IEC 27002 standard \cite{ISO2013ISO27002}. Research developments include proposals of a socio-technical threat model for software supply chains \cite{wang2013socio} and a taxonomy of attacks against open-source components \cite{ladisa2022taxonomy}, as well as efforts to secure a specific software deployment pipeline \cite{bass2015securing}, to ensure that software is not compromised between stages in a pipeline \cite{torres2019toto}, to identify threats and countermeasures related to third-party packages and libraries \cite{barabanov2020systematics}, and to analyze malicious packages used in supply chain attacks \cite{ohm2020backstabber}.

We lack, however, a comprehensive view of the threats that affect software integrity in development pipe\-lines and of how these threats can be mitigated. This paper aims to bridge this gap. We introduce a generic, systematic threat model for software development pipelines, complemented by a discussion of suitable  mitigations. The threat model was developed using STRIDE~\cite{shostack2014threat}, a framework created by Microsoft that has gained wide adoption. A case study describing how the model can be adapted to a specific pipeline demonstrates its applicability. This case study shows that the adaptation is straightforward, and enabled us to identify all threats that were considered in the original work, as well as some others that had been overlooked. Our main contribution is the threat model introduced in Section~\ref{cap:modelagemAmeaca}. Other contributions include:
\begin{itemize}
\item A contemporary review of more than 25 attacks against software development pipelines that have been discussed in the literature (Section~\ref{sec:pipeline});
\item A new classification of the consequences of threats to software development pipe\-lines (Section~\ref{sec:threat-consequences}); and
\item A case study applying our model to a pipeline from the literature, showing how the model can be adapted to specific implementations with limited effort (Section~\ref{cap:modelagemBass}).
\end{itemize}

Given our concern with the introduction of vulnerabilities in software by exploiting its development pipeline, our threat model considers only threats to integrity, not to confidentiality or availability. For the sake of generality, we consider threats at the design level; this means that vulnerabilities in specific tools are out of scope for this paper. Hardware attacks are also outside the scope. Such attacks are less prevalent, and, even though there has been a report about the insertion of malicious chips on server motherboards \cite{robertsonmichaelriley2018}, it has not been substantiated thus far \cite{owen2018apple, gallagher2019appleinsider}. Therefore, we limit our discussion to software issues. 

The remainder of the paper is organized as follows. Section~\ref{cap:revisaoLiteratura} presents a reference pipeline from the literature along with an up-to-date review of real-world attacks against each stage of the pipeline, and summarizes the STRIDE threat modeling framework. Section~\ref{sec:trabalhosRelacionados} reviews related work. Section~\ref{cap:modelagemAmeaca} introduces our threat model for software development pipelines and discusses how the threats can be mitigated. Section~\ref{cap:modelagemBass} presents the case study. Section~\ref{cap:conclusao} concludes the paper.

%% file: chapters/2_Background.tex
\section{Background} \label{cap:revisaoLiteratura}

This section reviews two main issues related to this paper. Section~\ref{sec:pipeline} presents software development pipelines, discussing some security risks and surveying documented attacks associated with each step, and Section~\ref{sec:modelagem} reviews threat modeling using the STRIDE framework.

\subsection{Software development pipeline} \label{sec:pipeline}

\begin{figure}
    \centering
    \caption{Typical software development pipeline}
    \includegraphics[width=1.0\linewidth]{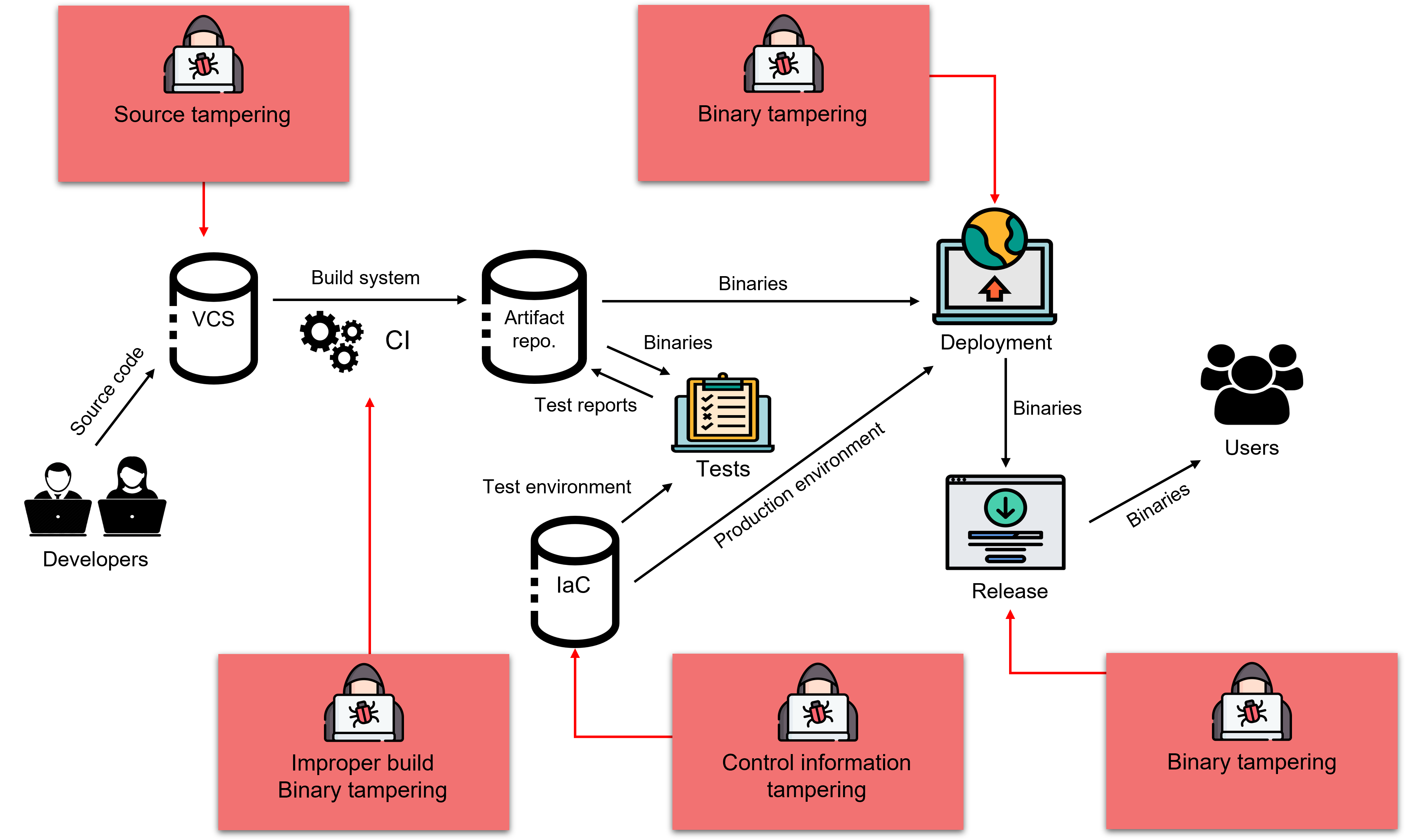}
    \label{fig:pipeline}
\end{figure}

Turning source code written by developers into software products that are made available to end users is a complex process involving several activities, ranging from building to distributing or releasing software packages and updates. This complex process is implemented using a software development pipeline comprising multiple stages, with no rigid definition for these stages. Figure~\ref{fig:pipeline} depicts a typical pipeline comprising five stages \cite{adams2016modern}: Integration: branching and merging, Continuous Integration: building and testing, Infrastructure as Code, Deployment, and Release. Although not all projects have a pipeline identical to the one in Figure~\ref{fig:pipeline}, adopting this pipeline as a reference allows all activities in the path between developers and users to be included in our threat model. Projects with simpler pipelines may disregard parts of the threat model, and projects with more complex ones may use our model as a baseline. The stages of the reference pipeline are detailed below.

\subsubsection{Integration: branching and merging} \label{sub:integracao}

The first stage of the pipeline involves source code changes. Typically, a developer makes code changes in a private source tree, then propagates those changes to the source tree used by her team. Finally, the developer sends these changes to the master source tree for the project \cite{adams2016modern}. To help in this process, many software organizations use version control systems (VCS) such as Subversion or Git. With version control, the developer creates a new branch where she can make the necessary changes. After making the desired changes, the developer can merge her branch with the master branch, making the changes visible to other team members. Another way to deal with branches and merges involves pull requests, which are typical of open source projects \cite{bit2021pull}. In this model, a developer clones a repository (even if he only has read access) and makes modifications to a branch in his private repository. When this developer wants to integrate his changes to the original repository, he submits a pull request indicating the branch where the modifications are. This pull request is reviewed by a developer with write access to the original repository, who may accept the request, decline it, or ask for changes before acceptance. The VCS will log the approvals of changes to the master branch of the project.

Attacks in the Integration stage involve unauthorized modifications to source code. Examples of real attacks include:
\begin{itemize}
    \item an attempt to modify the Linux kernel to insert a vulnerability \cite{corbet2003:backdoor};
    \item the case where attackers inserted unauthorized code into the operating system of Juniper network equipment, creating a backdoor for remote access to the equipment and allowing it to monitor and decrypt traffic \cite{Juniper2015};
    \item the case where Chrome developers had their credentials stolen and attackers were able to modify browser extensions, compromising millions of users \cite{maunder2017psa}; 
    \item a case of credential theft that happened in the OpenDev collaborative development project, where the compromise of an account with administrative privileges in a code review tool allowed for tampering with the source code repositories hosted by OpenDev, among other damages \cite{sharwood2020};
    \item the case where an attacker compromised a developer’s account and inserted a backdoor into Ruby Gems packages, enabling cryptocurrency mining and remote code execution on infected servers \cite{costa2019, goodin2019supplychain}.
\end{itemize}
Developers may also fall victim to compromised development tools (e.g., an integrated development environment, a package manager). Gerste  \cite{gerste2022securing} reports on vulnerabilities found in some popular package managers for languages such as PHP, Python, and Ruby. If an attacker managed to trick a developer into opening malicious packages using a compromised package manager, he would be able to steal and/or insert vulnerabilities into source code.

\subsubsection{Continuous Integration: building and testing} \label{sub:ic}

Continuous Integration (CI) refers to continually polling the VCS for new commits or merges, identifying those revisions, compiling them, and running an initial set of tests to verify that the changes have not caused any problems with the project \cite{adams2016modern}. To help in these activities, we can leverage CI tools such as Jenkins \cite{jenkins2016} or similar.

This stage includes both automated and manual testing. According to \cite{humble2010continuous}, the role of a tester is to (manually) check that software meets its acceptance criteria. Testers are not allowed to modify source code or binary files. All testing stages retrieve products built by the CI process and stored in an artifact repository \cite{adams2016modern}. One of the most significant components of the Continuous Integration step is the build system, which generates project outputs as binaries, libraries, or packages from source code \cite{adams2016modern}.

Attacks in the Continuous Integration stage involve the insertion of vulnerabilities into the software by corrupted tools, packages, and libraries. Regarding the build system, we can mention the classic Trojan compiler described by Ken Thompson \cite{thompson1984reflections}; this was an experiment under controlled conditions, but there are also cases where users were actually harmed. One example is XcodeGhost, a counterfeit version of Apple's Xcode development environment that included malicious code alongside the actual app code \cite{claudxiao2015}. This attack was responsible for over 4,000 infected apps in the iOS App Store \cite{lorenzofranceschibicchierai2021}. Another example is the case where the Webmin tool was compromised \cite{webmin2019}. Attackers broke into a build server that had a local copy of the source code and introduced vulnerabilities into a file without changing its modification timestamp. This modified version was used for builds that did not pull the source from the VCS and were distributed on SourceForge, affecting an undisclosed number of users. 

Modern software relies heavily on external packages and libraries, making these components attractive targets.
Examples of real-world cases include the use of typosquatting to register packages with malicious content in the npm and PyPI repositories using similar names to those of legitimate packages, to infect developers who made typos  \cite{catalincimpanu2017, sharma2022string, sharma2022python}. 
In 2021, a newly discovered attack, called dependency confusion, made it possible to breach the internal systems of more than 35 major companies, including Microsoft, Apple, PayPal, Shopify, Netflix, Yelp, Tesla, and Uber \cite{birsan2021dependency}. According to \cite{birsan2021dependency}, who disclosed the attack, if a dependency package used by an application exists both in an open-source repository and in its private build, the public version takes precedence. The attack consists of creating fake projects in open-source repositories such as npm, PyPI, and RubyGems, with the same names used in private projects (inferred from source code found in public repositories such as GitHub) and with malicious content. 
%
Since then, other malicious packages have been identified, 34~in the npm repository \cite{polkovnychenko2021malicious, goodin2021maliciousNPM, sharma2022john}, and 1275 in the PyPI repository \cite{sharma2022pypi}. The packages had different infection strategies, including dependency confusion, typosquatting, and Trojan functionality.

Package versions may also introduce issues. For instance, it is possible to have two distinct files X and Y with the same package name but with different versions, which may even be signed. The desired package is file X, but file Y is used at build time. A real case example is the dependency incident with the \emph{left-pad} package, in which a developer removed the \textit{left-pad} package from npm, causing a lot of problems for other developers \cite{gallagher2016}. To solve the problem, another developer replaced the deleted package with one that was functionally equivalent. In this replacement, a malicious developer could have inserted some malicious code into the package and harmed the software dependent on that package. This incident with the \textit{left-pad} package can also be considered a version issue.

\subsubsection{Infrastructure as Code} \label{sub:infraestrutura}

We can deploy a new version of the system for testing or production to a server, cloud, container, or virtual machine \cite{adams2016modern}. In this stage, we can use the term ``infrastructure'' or  ``environment'' to refer to these technologies. Infrastructure as Code (IaC) automates environment setup, generating the infrastructure based on specifications developed in a domain-specific programming language such as Puppet, Chef, Ansible, or similar \cite{morris2016infrastructure}. These tools automate the provisioning of virtual infrastructures and the installation and configuration of operating systems and ancillary services, ensuring that applications have a consistent and correct environment to run on.

Vulnerable specifications will generate vulnerable infrastructure, which is particularly dangerous when it comes to production environments. Scripts with vulnerable specifications may, for instance, deploy a vulnerable OS or backdoored system services to facilitate later intrusions, or create architectural elements that force network traffic to pass through a proxy that can observe and modify the traffic, thus enabling man-in-the-middle (MITM) attacks. A recent report identified nearly 200,000 vulnerabilities in IaC specifications \cite{PaloAlto2020}. Rahman and colleagues \cite{rahman2019seven} identified 21,201 occurrences of seven security smells (coding patterns that indicate weaknesses that may lead to security violations) in Puppet scripts. The same authors replicated the study using 50,323 Ansible and Chef scripts, and found 46,600 security smells \cite{rahman2021security}.

\subsubsection{Deployment} \label{sub:implantacao}

The Deployment phase prepares the products for Release. For example, for web applications, Deployment might consist of copying a set of files over the network to the correct directory on a web server \cite{adams2016modern}. In many cases, the software remains idle between the Deployment and Release steps.

Attacks at the Deployment stage involve unauthorized modifications of executable code. An example of a real attack is the spread of NotPetya malware through tampered updates to accounting software \cite{wheeler2018:swa}. Another example is the insertion into the third-party package repository used by Python developers (PyPI) of tampered libraries with the same name as the official ones (which should not be installed using PyPI) \cite{goodin}.

\subsubsection{Release} \label{sub:lancamento}

In this final stage of the pipeline, the deployed products are released and made available to end users. There are several ways to do so, such as offering a new app or version on an app store, releasing a web application, and making binaries available for download on a website. Attacks at the Release stage involve replacing or tampering with executables and components made available to users. Examples of real attacks include inserting malicious code into Microsoft Windows (version not specified) downloaded via Tor \cite{hern}. Another example is breaking into Avast's software development or distribution process at some stage before digitally signing the binaries. This attack allowed malware-infected versions of the CCleaner tool to be distributed to more than two million users \cite{tomwarren2017, brumaghin2017}.

\subsection{STRIDE Threat Modeling} \label{sec:modelagem}

A threat can be defined as a potential for violation of security \cite{rfc4949}. Threat modeling is a process for identifying, documenting, and mitigating threats to a system \cite{shostack2014threat}. Correct identification of threats to the system, and adequate mitigations for them, allow reducing the possibility of potential attackers being successful \cite{myagmar2005threat}. Because of this, threat modeling considers the system from the attackers' point of view, allowing us to anticipate where attacks may occur, making it possible to define \textit{what} the system was designed to protect and from \textit{whom} \cite{myagmar2005threat}.

There are many threat modeling methods. Twelve of them are analyzed in \cite{shevchenko2018threat}, 
which concludes that STRIDE (developed by Microsoft) is the most mature among them, helps identify relevant mitigation techniques, and is easy to use. STRIDE is an acronym that stands for the following six classes of threats \cite{shostack2014threat}:

\begin{itemize}
   \item Spoofing: pretending to be something or someone you are not;
   \item Tampering: corrupting or tampering with data;
   \item Repudiation: denying having done something;
   \item Information disclosure: revealing information to unauthorized users;
   \item Denial of service: interrupting the provision of services to legitimate users; and
   \item Elevation of privilege: when a user or program can act on the system with privileges beyond those granted to it.
 \end{itemize}

STRIDE threat modeling starts from a system model using a data flow diagram (DFD). A DFD is a classic software engineering modeling tool \cite{le2000understanding} that represents how data moves across a system using the following components:

\begin{itemize}
   \item External entities: elements that receive and send data to/from the system but are outside its scope;
   \item Processes: components that manipulate input data to generate output data, transforming or redirecting the data;
   \item Data stores: containers where processes store data that can be retrieved by the same or another process, or by an external entity; and
   \item Data flows: represent data movement between an entity, a process, and/or a data store.
 \end{itemize}
Trust boundaries indicate components that belong to different security contexts (e.g., protection domains), and are added to the diagram to help in identifying threats \cite{shostack2014threat}.

In STRIDE threat modeling, DFD components are analyzed for their susceptibility to threats of each class \cite{hernan2006threat}. There are several approaches to structuring this analysis, including STRIDE-per-element, STRIDE-per-interaction, DESIST, and the Elevation of Privilege game \cite{shostack2014threat}. We adopt STRIDE-per-element because it is more prescriptive, making it easier to find threats by focusing on a set of threats against each element, and performed better than STRIDE-per-interaction in a controlled study \cite{tuma2018two}.

In STRIDE-per-element, for each element of the DFD (external entity, process, data flow, and data store), we consider which threats this element is subject to, according to the possibilities contained in Table~\ref{tab:stride}. This table shows, for example, that external entities are only subject to spoofing and repudiation threats, and it is not necessary to consider the other classes. In the case of data stores, the threat of repudiation applies only to logs (thus the `?').

\begin{table}
\centering
\caption{STRIDE-per-element}
\label{tab:stride}
\begin{tabular}{lllllll}
\hline
                       & S & T & R & I & D & E \\ \hline
External entity       & \checkmark &   & \checkmark &   &   &   \\ 
Process              & \checkmark & \checkmark & \checkmark & \checkmark & \checkmark & \checkmark \\ 
Data flow         &   & \checkmark &   & \checkmark & \checkmark &   \\ 
Data store &   & \checkmark & ? & \checkmark & \checkmark &   \\ 
\end{tabular}
\end{table}

%% file: chapters/3_RelatedWork.tex
\section{Related work} \label{sec:trabalhosRelacionados}

In this section, we discuss previous work on the security of software development pipelines and software supply chains.

Levy \cite{levy2003:poisoning} was one of the first to address the security of software development pipelines. He observed how the shift from software built in closed environments and distributed on physical media towards a network-based environment with multiple builders (such as open-source operating systems projects), who obtain source code and distribute binary code over the Internet, created more opportunities for compromising software. He also reviewed some attacks on open-source software between 2001 and 2002 that involved introducing vulnerabilities in source code and compromising sites used for distribution. The paper discussed some threats to development pipelines together with applicable mitigations, without aiming at a holistic view of the issue.

In \citeonline{simpson2010software}, the authors discuss software supply chain security, focusing specifically on integrity practices and controls. They present three goals for software assurance: security (avoiding vulnerabilities in source code, designing with security in mind), integrity (ensuring that code is built and delivered correctly to users), and authenticity (ensuring that counterfeit software can be identified). The report presents a list of software assurance mechanisms adopted in the industry, featuring both technical controls and contractual requirements, and considering all three goals. The technical controls presented can be summarized as proper physical and logical access control at all levels, security testing, mechanisms for securing delivery (malware scanning, code signing) and for ensuring authenticity (encrypted components, notifying users about unauthentic software and preventing its execution). The report, however, lacks an explicit list of the threats considered.

A socio-technical threat model for software supply chains is introduced in \cite{wang2013socio,al2015socio}. The threat model features 30~social and 24~technical threats, with slight variations (e.g., whether a vulnerability is accidental or intentional) counted separately. Four social and eight technical threats refer to integrity, while the others refer to confidentiality or availability. Interviews with four experts validated the model. The threats to integrity are discussed in a more general fashion than in our model, and only a part of them are mapped to stages in a development pipeline; nevertheless, our model includes all threats described in \cite{wang2013socio}. Countermeasures are discussed in passing, while real-world cases are not considered. 

Shaw \cite{shaw} presented a list of publicly disclosed attacks against the software development pipeline that occurred between 2014 and 2017. The author associates the attacks with specific steps in the development pipeline, i.e., whether each attack occurred in the development tools, source code, distribution, or software updates. He also shows the growing number of attacks on the software development pipeline. Although the author has discussed several attacks, he has not presented countermeasures for these attacks or a threat model consolidating them. 

A report from ENISA \cite{enisa2021threat} proposes a taxonomy for supply chain attacks that classifies attacks according to the techniques used to compromise suppliers and customers, and the supplier and customer assets targeted. The report takes a broader view of what constitutes a supply chain attack: in addition to code, libraries, and configurations (the main assets that comprise software), supplier assets that may be targeted also include data, hardware, people, and processes. The document does not attempt to enumerate threats and discusses mitigations only superficially.

Barabanov and colleagues \citeonline{barabanov2018current} develop an attacker model and propose a systematic list of information security threats in software development processes. They consider both intentional and unintentional threats against confidentiality, integrity, and availability. The list was developed using attack trees and features 35~threats, but the paper actually describes only one of them, the introduction of vulnerabilities through compromised third-party components. The paper does not discuss either actual attacks or countermeasures. Their work is further developed in \cite{barabanov2020systematics}, which first provides a high-level overview, based on the literature, of threats to the software development pipeline and of measures to mitigate them. It then zooms in on attacks via open-source packages and libraries, providing a more detailed threat model, discussing some real-life attacks, and proposing countermeasures. While on the surface the scope of the work is wide, specifics are limited to inserting vulnerabilities via compromised third-party packages and libraries.

He and colleagues \citeonline{he2015model} have as main concern the risks to the software and hardware supply chains in the context of information technology acquisitions. They propose a model for the product development lifecycles of software and hardware commercial-of-the-shelf (COTS) products, and outline threats to the various phases of the lifecycle, along with some countermeasures.
While the report points out that the development and distribution phases hold the largest risk for the software supply chain, it does not discuss the security of the development pipeline in depth.

A group at MITRE \citeonline{miller2013supply,reed2014supply} developed a catalog of 41~supply chain attack patterns that involve the malicious insertion of hardware, software, firmware, or system data; 15~of those patterns apply to software development. In addition to the catalog of attacks, they also provide a list of 20~recommended countermeasures, of which~13 are software-focused. Not all countermeasures are technical (\emph{e.g.}, some deal with personnel management), and the correspondence between attacks and countermeasures is not explicitly defined (\emph{i.e.}, they do not say which attacks a countermeasure is intended to neutralize). As this catalog was developed in the context of the US Department of Defense (DoD) acquisitions, it is broader in scope and less specific than the threats and countermeasures advanced in our work. Moreover, the reports do not discuss real-world cases directly.

Bass and colleagues \citeonline{bass2015securing} discuss how a deployment pipeline can be subverted, presenting three scenarios. The first is when the deployed image is not valid, the second is when an image is deployed without going through the entire pipeline, and the third is when the production environment is accessible from a different environment. The paper focuses only on the first scenario. The authors introduce an engineering process based on trusted components to secure the pipeline. This process restricts access to the components and reduces permissions, so an attacker can only reach trusted components. The paper does not address real-world cases and discusses threats and countermeasures for a specific pipeline involving Chef, Jenkins, Docker, Github, and AWS, being narrower in focus than our work.

Paule \citeonline{paule2018securing} presents existing approaches and methods for detecting vulnerabilities in continuous delivery pipelines. The threat modeling method used was STRIDE. The thesis presents a review of vulnerability detection tools and a case study in a company which applied selected tools to verify the security level in two industrial continuous delivery pipelines. The focus of the work is the detection of vulnerabilities, and it does not include protection mechanisms.

Software assurance (SwA) tools are static and dynamic vulnerability detection tools; the former analyze source code, while the latter analyze running code. We can use these tools in the Integration, Continuous Integration, and Infrastructure as Code stages to detect vulnerabilities in software during development. SwA tools may pose a risk because they have privileged access to information such as source code, which may prompt an attacker to create a malicious tool. An attacker can even exploit unintended vulnerabilities of a tool. In \citeonline{wheeler2018:swa}, the authors identified practical approaches to protect development pipelines from the risks that can be caused by SwA tools. The authors also discussed the need to adopt a set of these tools to detect vulnerabilities with sufficient coverage to obtain software security assurance.

Torres-Arias and colleagues \citeonline{torres2019toto} introduce \textit{in-toto}, a framework that aims to (cryptographically) guarantee the integrity of the software supply chain. The idea of this tool is to protect, through encryption, the products generated at each stage of the supply chain,  allowing each phase to receive legitimate data. This way, it is also possible to verify which stages (of the chain) the software went through. However, the work does not take into account compromised steps, it only guarantees the integrity of the flow between the phases, i.e., if the products were not corrupted during the journey between one step and another. Therefore, if a step is compromised, it can generate vulnerable products, and these will serve as input to the next step in the pipeline.

Ohm and colleagues \cite{ohm2020backstabber} analyze 174~malicious Node.js, Python, and Ruby packages used in supply chain attacks. They also introduce two attack trees, one for injecting malicious code into dependency trees and the other for activating the the injected logic in dependent software, and discuss some countermeasures for malicious packages. This work is continued in \cite{ohm2022feasibility}, where they examine the feasibility of using supervised machine learning to automate the detection of malicious packages based on the dataset from \cite{ohm2020backstabber}. The attack tree for malicious code injection is the aspect closest to our work, and our threat model considers all attacks contained therein.

Ladisa and colleagues \cite{ladisa2022taxonomy} propose a taxonomy of attacks against open-source components with the aim of compromising software that includes such components. They present an attack tree representing different ways of tricking developers into using malicious components and of compromising the original ones by subverting their development pipelines. They also discuss countermeasures and real-world cases. The attack tree and the countermeasures were validated and assessed by domain experts and software developers. Given the focus on open-source components used by other software projects, they do not consider hosted applications. All attacks from the taxonomy are present in our threat model, but are generally discussed at a finer-grained level in \cite{ladisa2022taxonomy}.

Table~\ref{tab:trab_rel} summarizes how related work addresses threats to the stages of our reference pipeline. In each cell, ``T'' means that the reference discusses threats for the corresponding stage, and ``C'' indicates that it discusses countermeasures; we do not categorize references according to their depth of treatment, which varies. None of the references addresses both threats and countermeasures encompassing all stages of the development pipeline, which is the main differential of the present work.

\begin{table}[htbp]
\centering
\caption{Summary of related work}
\label{tab:trab_rel}
\def\arraystretch{1.8}
\begin{tabular}{lccccc}
\hline
Reference & Integration & CI & IaC & Deployment & Release \\ \hline
\citeonline{barabanov2020systematics} & [C] & [T, C] & -- & [C] & [C]  \\ 
\citeonline{barabanov2018current} & [T] & [T] & -- & [T] & [T]  \\ 
\citeonline{bass2015securing} & [C] & [T, C]  & [T, C] & [C] & --  \\ 
\citeonline{enisa2021threat} & [T] & [T]  & -- & [T] & [T],  \\ 
\citeonline{he2015model} & [T] & [T, C] & -- & -- & [T]  \\ 
\citeonline{ladisa2022taxonomy} & [T, C]  & [T, C] & -- & -- & [T, C]  \\ 
\citeonline{levy2003:poisoning} & [T, C]  & [T, C] & -- & [T, C] & [T, C]  \\ 
\citeonline{miller2013supply, reed2014supply} & [T, C] & [T, C] & [T] & [T, C] & [T, C]  \\ 
\citeonline{ohm2020backstabber} & [T, C]  & [T, C] & -- & [T, C] & [T, C]  \\ 
\citeonline{paule2018securing} & [T] & [T] & [T] & [T] & [T]  \\ 
\citeonline{shaw} & [T]  & [T] & --  & [T] & [T] \\ 
\citeonline{simpson2010software} & [C]  & [C] & [C] & [C] & [C]  \\ 
\citeonline{torres2019toto} & [T, C]  & [T, C] & -- & [T, C] & [T, C]  \\ 
\citeonline{wang2013socio} & [T]  & [T] & -- & [T] & [T]  \\ 
\citeonline{wheeler2018:swa} & [T, C] & [T, C] & -- & [T, C] & [T]  \\ 

\end{tabular}
\end{table}

There are several approaches to finding threats to the security of the software development pipeline, one of which is STRIDE\@. STRIDE has been used to develop threat models in various contexts \cite{abomhara2015:stride, Cagnazzo2018:threatmod, jelacic2017stride, karahasanovic2017:adapting, Khan2017:stride, ma2016:threat, Marksteiner2019:cybersec, mockel2010threat, sanfilippo2019, sattar2021, schmittner2019:threat, Wilhjelm2020}, being a well-tested, established approach.

%% file: chapters/4_ThreatModeling.tex
\section{Threat Modeling the Software Development Pipeline} \label{cap:modelagemAmeaca}

This section introduces our threat model for the software development pipeline. Section~\ref{sec:modelagemPipeline} presents a DFD model of the pipeline. Section~\ref{sec:modelagemAmeacas} details the threat model and discusses possible mitigations for the threats found.

\subsection{Modeling the development pipeline} \label{sec:modelagemPipeline}

\begin{figure}
    \centering
    \caption{DFD for the software development pipeline}
    \includegraphics[width=1.4\linewidth, angle = 90]{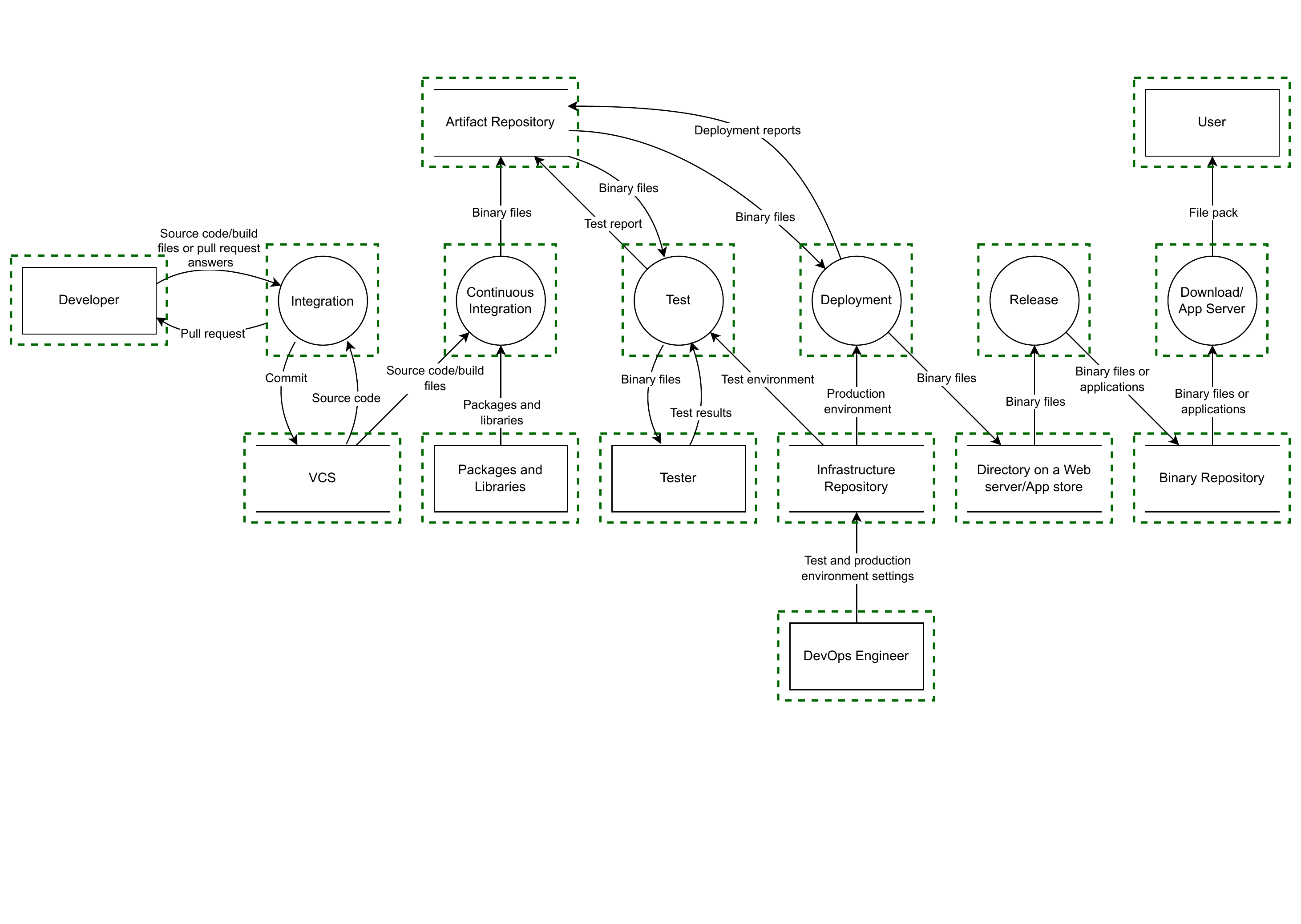}
    \label{fig:DFD_Pipeline}
\end{figure}

To develop a threat model for the software development pipeline presented in~Section~\ref{sec:pipeline}, we had first to model the pipeline using a DFD, shown in Figure~\ref{fig:DFD_Pipeline}. 
In the diagram, going from left to right we see a \fontDFD{Developer} accessing the first phase of the pipeline, which is represented by the \fontDFD{Integration} process. Here, the \fontDFD{Developer} sends the source code and the build files, which are stored in the \fontDFD{VCS} repository and retrieved by the \fontDFD{Integration} process when needed. A senior \fontDFD{Developer} also receives pull requests and sends accept/reject decisions to the \fontDFD{Integration} process. At the end of this phase, the source code and the build files are sent to the \fontDFD{Continuous Integration} process. This process can incorporate packages and libraries from an external entity named \fontDFD{Packages and Libraries}, which is outside the organization's control as these are third--party packages and libraries that will be used to help the development of the software. Completing the \fontDFD{Continuous Integration} phase, the binary files are sent to the \fontDFD{Artifact Repository}, and they can be used by the \fontDFD{Test} and \fontDFD{Deployment} processes. 
The \fontDFD{Test} process involves automated tests. Manual tests are carried out by the \fontDFD{Tester} external entity, which receives binary files as input and returns test results. 

\fontDFD{DevOps Engineer} is an external entity responsible for configuring test and production environments using IaC scripts. These scripts are stored in the \fontDFD{Infrastructure Repository}, from where they are retrieved by the \fontDFD{Test} and \fontDFD{Deployment} processes (for test and production environments respectively). The \fontDFD{Deployment} phase ends with  binary files being sent to a \fontDFD{Directory on a Web server} or to an \fontDFD{App store}, depending on the project. 
These binary files are then sent to the \fontDFD{Release} process, which stores them on a \fontDFD{Binary Repository}. The \fontDFD{Download/App Server} provides a \fontDFD{User} with access to  downloadable files or hosted applications, depending on the environment.

To make the analysis more general, each element of the diagram is considered to be in its own trust boundary (delimited by a dashed line). When there is more than one element within the same boundary (for example, two processes running on the same system), some threats become meaningless, and it becomes unnecessary to mitigate them \cite{Dotson2019:pcs}. For example, as processes within a boundary trust each other, there is no need to take into account the threat of process spoofing. Thus, when a pipeline implementation places multiple elements within the same trust boundary, some threats identified in the model presented in Section~\ref{sec:modelagemAmeacas} (notably those that are a consequence of mutual distrust between elements) may be disregarded. 

\subsection{Threat model} \label{sec:modelagemAmeacas}

The threat model for the pipeline in Section~\ref{sec:pipeline} was developed by applying STRIDE-per-element to the DFD shown in Figure~\ref{fig:DFD_Pipeline}. To ease understanding, we first categorize the consequences of threats to the pipeline in Section~\ref{sec:threat-consequences}. Then, in Sections~\ref{sub:entidades} to~\ref{sub:depositosDados} we present the model and discuss mitigations for each threat. Since DFD elements of the same type feature many common threats, we devote a section for each element type. Given that our focus is on software integrity, we will not consider information disclosure and denial of service threats, as these involve confidentiality and availability, respectively. In Sections~\ref{sub:insiderThreats} and~\ref{sub:limitacoesTLS} we provide considerations on insider threats and the use of TLS, and in Section~\ref{sub:discussao3} we discuss our model.

\subsubsection{Threat consequences}
\label{sec:threat-consequences}

Essentially, a software development pipeline implements a process that transforms source code into binary code, which may be automatically deployed on computing infrastructure or made available for download to end users. Threats that result in malicious code must either subvert the input to this process, the output, and/or the process itself. We should also consider threats that do not produce malicious code but may indirectly affect integrity. The consequences of integrity-related threats to a development pipeline can thus be classified as:
\begin{enumerate}
    \item \emph{Source tampering}: enables an attacker to insert vulnerabilities and/or malicious logic into source code, which is correctly transformed into malicious binary code; 
    \item \emph{Binary tampering}: enables an attacker to insert vulnerabilities and/or malicious logic into binary code, even if it was correctly built from pristine sources, as well as include malware or malicious components (such as shared libraries) into installation images alongside correct binaries;
    \item \emph{Improper build}: enables an attacker to use malicious tools or other components (e.g., Trojan compilers, vulnerable/malicious libraries) to transform correct source code into malicious binary code;
    \item \emph{Control information tampering}: enables an attacker to change metadata to subvert the pipeline. Threats in this class do not produce malicious code by themselves, but they can allow compromised binaries to evade detection (e.g., by disabling tests or faking test results) or move along the pipeline; and
    \item \emph{Infrastructure tampering}: enables an attacker to subvert automatic code deployment. The threats in this class do not produce malicious code by themselves, but they purposefully weaken the infrastructure where correct code is deployed (e.g., by including backdoored services, disabling security controls, or using operating system versions with known vulnerabilities), to facilitate later attacks. 
\end{enumerate}
This classification allows us to discuss the impact of different threats to the pipeline in an easier to understand, more concise manner.

\subsubsection{External entities} \label{sub:entidades}

The DFD has five external entities: \fontDFD{Developer}, \fontDFD{Packages and Libraries}, \fontDFD{Tester}, \fontDFD{DevOps Engineer}, and \fontDFD{User}. The \fontDFD{Packages and Libraries} external entity provides third-party packages and libraries used to build the software. Given that this entity is outside the organization’s control and only provides components developed elsewhere, what matters is that the \emph{content} imported by the pipeline is legitimate (this threat is discussed in Section~\ref{sub:processos}). Whether or not the entity is legitimate (\textit{i.e.}, a possible spoofing threat) is irrelevant, because an illegitimate entity (a spoofed repository) can serve legitimate content, and a compromised legitimate entity can serve illegitimate content.

The threats found for \fontDFD{Developer}, \fontDFD{DevOps Engineer}, and \fontDFD{Tester} are:

\begin{itemize}
   \item \textit{Spoofing}: 
   \fontDFD{Developer}, \fontDFD{DevOps Engineer}, or \fontDFD{Tester} spoofing may lead to:
   \begin{itemize}
       \item Source tampering: if an attacker impersonates a \fontDFD{Developer} and inserts vulnerabilities into source code, or if an unprivileged \fontDFD{Developer} spoofs a privileged one and illegitimately approves pull requests. The latter case was not considered an elevation of privilege threat because the approvals are logged in the \fontDFD{VCS}. Thus, it becomes more viable for an attacker to steal the credentials of a user who already has approval privileges than for the attacker to approve a pull request on his behalf or on behalf of an unauthorized \fontDFD{Developer}, as this will make it clear in the \fontDFD{VCS} logs that there was  approval by an improper user. 
       \item Control information tampering: if an attacker spoofs a \fontDFD{DevOps Engineer} and alters test and production environments to legitimate tampered software, or impersonates a \fontDFD{Tester} and modifies test results to approve vulnerable software.
       \item Infrastructure tampering: if an attacker spoofs a \fontDFD{DevOps Engineer} and compromises the deployment infrastructure to facilitate later intrusions.
   \end{itemize}
   
   It is possible to mitigate user spoofing threats through strong authentication. The use of two-factor authentication (2FA) or multi-factor authentication (MFA) is recommended for sensitive accounts such as developer \cite{costa2019, goodin2019supplychain, maddox2018best, nist2019mfa, weinert2019password}.
   Another important mitigation is the management of user accounts, which involves processes to assign access permissions to resources according to the needs of each user, and grant/revoke access credentials to resources and systems \cite{TheisCommonSense2019}. Although 2FA/MFA is an important control, attention needs to be paid to the convenience and security of its implementation \cite{grassi2017nist,grimes2020hacking}: for instance, SMS-based 2FA remains popular despite some known weaknesses \cite{jover2020sms}, and phishing attacks are increasingly capable of bypassing multi-factor authentication \cite{michaelkan2019google, garethcorfield2022phishing, goodin2022phishing}.

   \item \textit{Repudiation}: a \fontDFD{Developer}, \fontDFD{DevOps Engineer}, or \fontDFD{Tester} may deny having sent malicious data to the system. For instance, a \fontDFD{Developer} may send source code with a backdoor to the \fontDFD{Integration} process (source tampering). Also, a \fontDFD{DevOps Engineer} may send a test environment that allows for legitimizing tampered software, or a vulnerable infrastructure for deployment (control information or infrastructure tampering). Finally, a \fontDFD{Tester} may submit fake test results that approve vulnerable software (control information tampering). While repudiation alone introduces no vulnerabilities, it may enable other threats, particularly user spoofing, to remain undetected. Repudiation threats can be mitigated by storing log data (recording who made changes and when those changes were made) \cite{kent2006:nist} as well as digitally signing commits \cite{vaidya2019commit,Git2022}.

 \end{itemize}
 
\subsubsection{Processes} \label{sub:processos}

The DFD has six processes: \fontDFD{Integration}, \fontDFD{Continuous Integration}, \fontDFD{Test}, \fontDFD{Deployment}, \fontDFD{Release}, and \fontDFD{Download/App Server}. These processes are subject to some common threats that can be mitigated in the same way (what changes are the consequences, which depend on the process):

\begin{itemize}
    \item \textit{Spoofing}: one threat against processes is server spoofing, that is, the possibility of an illegitimate process, data store, or external entity posing as the real one and supplying malicious data to the affected process. The possible consequences are:
    \begin{itemize}
        \item source tampering:
        \begin{itemize}
         
            \item via spoofed \fontDFD{Integration} (against \fontDFD{VCS});
            \item via spoofed \fontDFD{VCS} data store (against the \fontDFD{Integration} and \fontDFD{Continuous Integration} processes);
        \end{itemize}
        \item binary tampering:
        \begin{itemize}
            \item via spoofed \fontDFD{Continuous Integration} (against \fontDFD{Artifact Repository});            
            \item via spoofed \fontDFD{Artifact Repository} (against \fontDFD{Deployment});
            \item via spoofed \fontDFD{Deployment} (against \fontDFD{Directory on a Web server/App store});            
            \item via spoofed \fontDFD{Directory on a Web server/App store} (against \fontDFD{Release});
            \item via spoofed \fontDFD{Release} (against \fontDFD{Binary Repository});
            \item via spoofed \fontDFD{Binary Repository} (against \fontDFD{Download/App Server});
            \item via spoofed \fontDFD{Download/App Server} (against \fontDFD{User});
        \end{itemize}
        \item improper build:
        \begin{itemize}
            \item via spoofed \fontDFD{Packages and Libraries} (against \fontDFD{Continuous Integration});
            \item via spoofed \fontDFD{Continuous Integration} (against \fontDFD{Artifact Repository});
        \end{itemize}
        
        \item control information tampering:
        \begin{itemize}
            \item via spoofed \fontDFD{Test} (against \fontDFD{Tester} and \fontDFD{Artifact Repository});
            \item via spoofed \fontDFD{Artifact Repository} and \fontDFD{Infrastructure Repository} data stores (against \fontDFD{Test});
        \end{itemize}
        \item infrastructure tampering:
        \begin{itemize}
            \item via spoofed \fontDFD{Infrastructure Repository} (against \fontDFD{Deployment}).
        \end{itemize}
    \end{itemize}
    
    Server spoofing threats can be mitigated through server authentication, e.g.\ by using Transport Layer Security (TLS) certificates  \cite{rfc8446}  (Section~\ref{sub:limitacoesTLS} provides additional considerations on using TLS) or the SPIFFE framework \cite{spiffe2022}.

    \item \textit{Tampering}: the processes do not know whether the data they are receiving, whether from external entities, data stores, or other processes, is trustworthy. This threat affects all processes, with different consequences depending on the origin of the data and the receiving process. For instance, if  process \fontDFD{Integration} receives malicious data it may result in source tampering, while malicious input to \fontDFD{Continuous Integration} may result in source tampering, binary tampering, or improper build. It is possible to guarantee uncorrupted data using mechanisms such as permissions and digital signatures. Permissions ensure that only authorized principals (users or processes) can modify repository contents, while digital signatures guarantee that binaries have been created or certified by authorized principals. However, signatures only provide syntactic validation, that is, they establish a chain of custody, guaranteeing that the artifacts received by a process are the same that were produced by authorized principals. However, they are insufficient for semantic validation, that is, they do not guarantee that the software is correct and has not been victim of any kind of tampering or improper build. Software assurance tools that identify vulnerabilities in source or binary code \cite{kupsch2017bad, pistoia2007survey, wheeler2016soar} can be used as additional mitigation against tampering.

    Another threat is local falsification, i.e., a malicious process receives genuine data but writes falsified data to a store or sends it to an external entity (in this case, the process is an agent, not a victim). This also affects all processes, and may lead to any of the consequences defined in Section~\ref{sec:threat-consequences}, depending on the process; for instance, a malicious \fontDFD{Integration} process may perform source tampering, while a malicious \fontDFD{Deployment} process may cause binary and/or infrastructure tampering. This threat can be mitigated using intrusion tolerance techniques \cite{Verissimo2003}.
 
    \item \textit{Elevation of Privilege}: the spoofing threats to external entities presented in Section~\ref{sub:entidades} are equivalent to  elevation of privilege threats to the corresponding processes. Spoofing an external entity (\fontDFD{Developer}, \fontDFD{Tester},  \fontDFD{DevOps Engineer}) enables unauthorized access to system data and functions, which is the same consequence of a successful elevation of privilege in the processes. 
    
    Processes that do not interact with external entities are also susceptible to elevation of privilege, for instance through lateral movement \cite{purvine2016graph}. EoP may lead to any of the consequences defined in Section~\ref{sec:threat-consequences}, depending on the affected process. 
    Once a process is compromised, the other elements in the pipeline may receive falsified data and accept it as trusted.
    
    At the design level it is difficult to outline effective mitigations against EoP threats. Enforcing the principle of minimal privilege \cite{saltzer1975:protection} across processes (and their components, for non-monolithic processes) may limit the consequences of compromising any individual process or component, while intrusion tolerance techniques \cite{Verissimo2003} ensure correct operation even in the presence of a minority of compromised process replicas.

\end{itemize}

If a compromised process retains the proper authorizations and cryptographic keys, mitigations based on chain of custody (such as digital signatures) may no longer be effective, and vulnerability detection tools become even more important in identifying malicious artifacts.

In addition to these threats applicable to all processes, we found threats that are specific to the \fontDFD{Continuous Integration} process:

\begin{itemize}
    \item \textit{Tampering}: an attacker may insert a backdoor into a CI or build tool and thus introduce vulnerabilities into the software (resulting in an improper build). 
    To avoid this threat, it is the developer's responsibility to take due care when making use of third-party build tools. Tampered compilers can be mitigated using diversity, as in the diverse double compiling (DDC) technique \cite{wheeler2009fully, skrimstad2018improving}. Reproducible builds, a recent research topic \cite{lamb2022reproducible}, can also provide  mitigation for this problem.
    
    Another threat is that the \fontDFD{Continuous Integration} process does not know if the data received from \fontDFD{Packages and Libraries} is trustworthy and has not been corrupted. The ultimate cause may be a developer error, or the external entity (\fontDFD{Packages and Libraries}) may be compromised. If packages and libraries are corrupted then vulnerabilities can be inserted into the software via source tampering (if packages/libraries are in source form) or improper build (with binary packages/libraries). 
    The \fontDFD{Packages and Libraries} entity is outside the control of the organization, but some mitigations are available. Using cryptographic protection (e.g., TLS) for communication mitigates server spoofing and traffic tampering threats. Some vendors provide digitally signed packages, which make it harder for an attacker to tamper with the data. Another way is avoid trusting a single repository; for example, if the same version of a package is available in multiple official repositories, it is possible to compare files taken from several repositories to see if they are identical (if not, this indicates a possible compromise of the divergent repository). Ensuring that packages and libraries are free from vulnerabilities remains an elusive goal, however.

    An additional case that can be taken into account is the possibility of having two distinct files X and Y with the same package name but with different versions, which may even be signed. The desired package is file~X, however, at build time, file~Y is used \cite{gallagher2016}. 
    This incident can also be considered a versioning issue. 
    A similar threat is dependency confusion \cite{birsan2021dependency}. 
    These threats allow an attacker to achieve remote code execution and can then insert backdoors during builds. As mitigation, it is suggested that projects have a unique version identifier for each release \cite{thelinuxfoundation2020}.
    
\end{itemize}

\subsubsection{Data flows} \label{sub:fluxosDados}

The data flows in the DFD are susceptible to similar tampering threats, with consequences that differ slightly depending on the source and destination of each flow:

\begin{itemize}
    \item \textit{Tampering}: an attacker can alter data in transit (especially if components communicate over the network), with the following possible consequences:
    \begin{itemize}
        \item source tampering:
        \begin{itemize}
            \item any flow in the \fontDFD{Developer} $\to$ \fontDFD{Integration} $\to$ \fontDFD{VCS} $\to$ \fontDFD{Continuous Integration} path;
            \item \fontDFD{Packages/Libraries} $\to$ \fontDFD{Continuous Integration} (for packages/libraries in source form);
        \end{itemize}
        \item binary tampering:
        \begin{itemize}
            \item \fontDFD{Continuous Integration} $\to$ \fontDFD{Artifact Repository};
            \item any flow in the \fontDFD{Artifact Repository} $\to$ \fontDFD{Deployment} $\to$  \fontDFD{Directory on a Web server/App store} $\to$ \fontDFD{Release} $\to$ \fontDFD{Binary Repository} $\to$ \fontDFD{Download/App Server} $\to$ \fontDFD{User} path;
        \end{itemize}
        \item improper build:
        \begin{itemize}
            \item \fontDFD{Packages/Libraries} $\to$ \fontDFD{Continuous Integration} (for packages/libraries in binary form);
        \end{itemize}
        \item control information tampering:
        \begin{itemize}
            \item \fontDFD{Test} $\to$ \fontDFD{Artifact Repository};
            \item \fontDFD{Deployment} $\to$ \fontDFD{Artifact Repository};
            \item any flow in the \fontDFD{DevOps Engineer} $\to$ \fontDFD{Infrastructure Repository} $\to$ \fontDFD{Test} path (a tester may be induced to wrongly approve malicious binaries);
            \item \fontDFD{Test} $\to$ \fontDFD{Tester} (malicious binaries may be changed in a way that leads to positive feedback from testers);
            \item \fontDFD{Tester} $\to$ \fontDFD{Test} (negative feedback from testers may be changed to approve malicious binaries);
        \end{itemize}
        \item infrastructure tampering:
        \begin{itemize}
            \item any flow in the \fontDFD{DevOps Engineer} $\to$ \fontDFD{Infrastructure Repository} $\to$ \fontDFD{Deployment} path.
        \end{itemize}
    \end{itemize}
    Regardless of the consequence, the problem can be mitigated through cryptographic protection of communication channels, using, for instance, TLS (which, as described in Section~\ref{sub:entidades}, also provider server authentication and thus mitigates spoofing). 
    
    While external entities are deemed out of scope for a threat model \cite{hernan2006threat}, we note in passing that the \fontDFD{Developer} $\to$ \fontDFD{Integration} data flow may also be tampered with by a malicious development tool (e.g., an integrated development environment) that introduces vulnerabilities in the source code committed to a repository while presenting the correct version to the developer. The deceptive nature of this threat makes it difficult to mitigate at the source. Code reviews by other developers, or by the original developer using different tools (perhaps in another environment, such as a web interface), can be effective -- but these are process-related mitigations, not technical ones. Software assurance tools \cite{wheeler2018:swa, kupsch2017bad, pistoia2007survey} may also be used in the pipeline to detect vulnerabilities. The threat posed by malicious client tools also applies to the user interfaces employed by \fontDFD{Tester} (which may lead to control information tampering) and \fontDFD{DevOps Engineer} (which may lead to infrastructure or control information tampering). Developing a full threat model for the client side of a software pipeline is left as future work.

\end{itemize}

\subsubsection{Data stores} \label{sub:depositosDados}

There are five data stores in the DFD: \fontDFD{VCS}, \fontDFD{Artifact Repository}, \fontDFD{Infrastructure Repository}, \fontDFD{Directory on a Web server/App store}, and \fontDFD{Binary Repository}. Software integrity can only be compromised by tampering threats:

\begin{itemize}
 \item \textit{Tampering}: an attacker may tamper with data. 
The \fontDFD{VCS} repository stores source code, and its compromise can lead to source tampering. In the \fontDFD{Artifact Repository}, changes in test reports can make it possible to legitimate a tampered software (control information tampering), or binary files may be modified to call other functions or execute improper content (binary tampering). The \fontDFD{Infrastructure Repository} stores configuration files for test and deployment infrastructures, and changes to these files may be used to legitimate tampered software (control information tampering) or introduce vulnerabilities in the deployment infrastructure (infrastructure tampering). \fontDFD{Directory on a Web server/App store} and \fontDFD{Binary Repository} can lead to binary tampering if compromised. This is particularly dangerous: users are often unable to verify that the software they received is the software they wanted, \textit{i.e.}, that it is not malicious or fraudulent \citeonline{thelinuxfoundation2020}. Digital signatures may be used to certify that the software was built by authorized principals, but cannot guarantee that it is genuine. 

Mitigations against data store tampering include properly managing permissions and storing encrypted data. Software assurance tools may be used for detecting malicious software, but a supplier must take responsibility for scanning binaries it makes available for download or use over the network (in the case of hosted applications) and taking remedial action. However, if the development team chooses to make use of an \fontDFD{App store} instead of a \fontDFD{Directory on a Web server}, the responsibility for mitigating this threat rests entirely with the \fontDFD{App store} (the same applies to other threats).

\end{itemize}

\subsubsection{Insider threats} \label{sub:insiderThreats}

The hardware and software that comprise a development pipeline must be managed by one or more people, which are not represented in the DFD to avoid further complexity. Two types of threats must be considered: one is an attacker spoofing a system manager, and the other is a \emph{rogue} admin which turns out to be an attacker (an insider threat). By definition, administrative accounts can do (almost) anything on a system, and a user with administrative access is in a strong position to sabotage a pipeline. In fact, admin compromise can carry out all the threats listed in the previous sections (Sections~\ref{sub:entidades} to~\ref{sub:depositosDados}).

Dealing with insider threats is notoriously difficult \cite{hunker2011insiders}. Detecting and mitigating insiders encompass both managerial aspects (such as hiring and termination processes, awareness training, policies to improve job satisfaction, employee monitoring) and technical components, such as \cite{TheisCommonSense2019}: 
\begin{itemize}
    \item Strict password and account management policies and practices;
    \item Stringent access controls and monitoring policies on privileged users;
    \item Extensive monitoring and logging of administrative actions;
    \item Establishing a baseline of normal behavior for both networks and employees;
    \item Enforcing separation of duties and least privilege;
    \item Explicit security agreements for any cloud services, especially access restrictions and monitoring capabilities;
    \item Institutionalizing system change controls; and
    \item Secure backup and recovery processes.
\end{itemize}
Strong authentication is key to mitigate spoofing, but is powerless to deal with admins that choose to abuse their privileges to sabotage the pipeline. For this, separation of duties and least privilege are essential to restricting what any one admin can do (thus limiting potential damages), while extensive monitoring and logging play a vital role in detecting insider activity.

\subsubsection{Dealing with Limitations of TLS} \label{sub:limitacoesTLS}

TLS \cite{rfc8446} is the standard solution for data flow protection (providing confidentiality and integrity) and server authentication \cite{shostack2014threat}, and is advocated as a mitigation for several threats in the proposed modeling. TLS relies on a public-key infrastructure, and has known issues that fall into two broad groups: cryptographic vulnerabilities and trust model limitations \cite{clark2013sok}. With regard to cryptographic vulnerabilities, the best mitigation is to use the most current version of the protocol, which is currently TLS 1.3 \cite{rfc8446}, disabling previous versions.

As mentioned, TLS also has some limitations in its trust model. TLS clients (such as a web browser) typically have a set of certificate authorities (CAs) that are trust anchors, that is, certificates issued by one of these CAs (or another CA to whom one of the trusted CAs delegates the right to issue certificates) are accepted as legitimate. In addition, the trust model allows any CA to issue a certificate for any name, regardless of the will or consent of the responsible for that name. As a result, several scenarios can lead to the issuance of forged certificates that will be accepted as legitimate: the compromise of any trust anchor (\textit{i.e.}, any CA or private key), the fraudulent inclusion of a CA or key in the list of trust anchors used by a client, or some attack that convinces a genuine CA to issue an improper certificate. This allows for MITM attacks and server spoofing: for example, an attacker intercepts the data flow and presents a certificate as if it was the original server, causing the cryptographic protection to end in that interceptor and allowing the attacker to observe and rewrite the data. There are several attacks that enable data flow interception, such as ARP spoofing \cite{whalen2001:arp-spoofing}, DNS hijacking \cite{houser2021comprehensive}, and BGP hijacking \cite{nordstrom2004beware}. 

There are several proposals to work around the limitations of the trust model \cite{clark2013sok,diaz2019tls}. The most realistic are:

\begin{itemize}
    \item Narrow down the list of trust anchors to the necessary CAs. This is most feasible when the set of servers is under the control of the same organization, but may have continuity issues in the event of key revocation.

    \item Use pinning, which is the process of associating a server with one or more expected certificates or public keys \cite{owasp}. Two common ways of performing pinning are by loading the certificate or public key on the client beforehand, or trusting the first received certificate or public key (known as Trust on First Use, TOFU). Another form of pinning involves the use of DNS-Based Authentication of Named Entities (DANE) \cite{rfc6698,rfc7671}, which aims to associate domain names to certificates using DNSSEC\@. DANE allows domain owners to include information about the authentication credentials of their permanent services in their DNS records \cite{diaz2019tls}.

\end{itemize}

Therefore, when TLS is used to mitigate threats, it is recommended to adopt one of the aforementioned mechanisms (or another equivalent), to minimize the limitations of its trust model.

\subsubsection{Discussion} \label{sub:discussao3}

Table \ref{tab:modelagem} summarizes the threat model. The first column contains the elements. The second column indicates which threat type (STRIDE) the element is susceptible to. The third column describes this threat. Finally, the fourth column suggests mitigations. Table~\ref{tab:consequencias} provides a complementary view, summarizing the consequences of each threat found for its respective element in the DFD in Sections~\ref{sub:entidades} to~\ref{sub:depositosDados}. The first column contains the type of element. The second includes the elements, and the other columns have the threat consequences introduced in Section~\ref{sec:threat-consequences}.

The model contains 13 threats, with all stages of the software development pipeline affected. Most threats with immediate consequences involve providing malicious binaries (executables, libraries, install images), as seen in the columns for improper build and binary tampering in Table~\ref{tab:consequencias}. Even though there are many threats that may lead to control information tampering, they are only enablers to threats that affect sources, binaries, or deployment infrastructures. The pipeline stages affected by the highest number of threats are the \fontDFD{Continuous Integration} and \fontDFD{Deployment} processes; this does not imply that attention should be focused on these components, as some threats are easier to realize and/or to accomplish the results intended by an attacker than others. For instance, it is probably easier to replace a binary with a malicious one when it is at rest on a data store than while it is being sent over the network, or to introduce a vulnerability in source code than to implement a Trojan compiler that produces malicious binaries. In the same vein, mitigations differ in ease of implementation, cost, and effectiveness. For instance, adding TLS to existing HTTP connections between pipeline components is easy and inexpensive, while implementing intrusion-tolerant components is complex and costly. Some threats, such as receiving unreliable data, can only be partially mitigated. 

A threat model should first elicit all possible threats to a given system, and then decide how to deal with them \cite{shostack2014threat}. The likelihood and effectiveness of threats are considered in the latter step, not the former. While we include a list of mitigations for the threats in our model, deciding which ones will be implemented (if any) is up to those responsible for a given pipeline.  

Our threat model focuses on threats to integrity, considering threats to confidentiality and availability to be out of scope. It is thus, by design, incomplete. We note that even for integrity threats it is impossible to guarantee that the model is complete, as new attacks may be discovered in the future that are not addressed in the model. Even if this happens, it will remain a useful model that considers a broad swath of threats against software development pipelines.

\begin{table}
\centering
\caption{Summary of threats and mitigations found for the DFD}
\label{tab:modelagem}
\def\arraystretch{1.3}
\scalebox{0.85}{
    \begin{tabular}{|p{4cm}|p{2cm}|p{3.5cm}|p{4.5cm}|p{1cm}}
    \cline{1-4}
    \multicolumn{1}{|c|}{\textbf{DFD element}} & \multicolumn{1}{c|}{\textbf{Threat type}} & \multicolumn{1}{c|}{\textbf{Threat}} & \multicolumn{1}{c|}{\textbf{Mitigation}} &  \\ \cline{1-4}
    \multirow{3}{*}{\begin{minipage}{4cm}\fontDFD{Developer}\\\fontDFD{DevOps Engineer}\\\fontDFD{Tester}\end{minipage}} & Spoofing & User spoofing & Authentication &  \\ \cline{2-4} &  Repudiation & Deny sending data to the system  & Logging, commit signing &  \\ \cline{1-4}
    \multirow{6}{*}{\begin{minipage}{4cm}\fontDFD{Integration}\\\fontDFD{Continuous Integration}\\\fontDFD{Test}\\\fontDFD{Deployment}\\\fontDFD{Release}\\\fontDFD{Download/App Server}\end{minipage}} & Spoofing & Server spoofing & TLS certificates &  \\ \cline{2-4} & 
        \multirow{3}{*}{Tampering} & \raggedright Receiving unreliable data & Permissions, digital signatures, tools for detecting vulnerabilities in source or binary code &  \\ \cline{3-4} &  & Local spoofing & Intrusion tolerance techniques &  \\ \cline{2-4} &
        \multirow{2}{*}{\begin{minipage}{2cm}\raggedright Elevation of privilege\end{minipage}} & \raggedright Unauthorized access to system data & \raggedright Authentication, minimal privilege, intrusion tolerance &  \\ \cline{3-4} &  & \raggedright Unauthorized access to entity functionality & \raggedright Authentication, minimal privilege, intrusion tolerance &  \\ \cline{1-4}
    \multirow{5}{*}{\fontDFD{Continuous Integration}} & \multirow{5}{*}{Tampering} & Subverted tools & \raggedright Developer prudence, tool diversity, reproducible builds &  \\ \cline{3-4} & & \raggedright Receiving unreliable data & \raggedright HTTPS connections, digital signatures,  repository diversity &  \\ \cline{3-4} & & \raggedright Packages with the same name & \raggedright Unique version identifiers for each release & \\ \cline{1-4}
    \multirow{1}{*}{\textit{Data flows (all)}} & Tampering & \raggedright Altering data during communication & TLS cryptography or equivalent &  \\ \cline{1-4} 
    \multirow{1}{*}{\begin{minipage}{4cm}\vspace*{0.4em}\fontDFD{Developer} $\to$ \fontDFD{Integration}\\\fontDFD{Tester} $\to$ \fontDFD{Test}\\\fontDFD{DevOps Engineer} $\to$\\ \hspace*{1em} \fontDFD{Infrastructure Repository}\end{minipage}} & \begin{minipage}[t][1.65cm][t]{2cm}Tampering\end{minipage} & \raggedright Malicious development tool & Software assurance tools &  \\ \cline{1-4} 
    \multirow{1}{*}{\begin{minipage}{4cm}\raggedright\vspace*{0.4em}\fontDFD{VCS}\\ \fontDFD{Artifact Repository}\\\fontDFD{Infrastructure Repository}\\\fontDFD{Web server/App store}\\\fontDFD{Binary Repository}\end{minipage}} & \begin{minipage}[t][2cm][t]{2cm}Tampering \end{minipage}& Improper data alteration & \raggedright Permission management,\\data-at-rest encryption &  \\ \cline{1-4}  
\end{tabular}

}
\end{table}

\begin{table}
\centering
\caption{Summary of threat consequences found for the DFD}
\label{tab:consequencias}
\def\arraystretch{1.3}
\scalebox{0.85}{
\begin{threeparttable}[b]
\begin{tabular}{|p{1.3cm}|p{1.6cm}|p{2cm}|p{2cm}|p{2cm}|p{2cm}|p{2cm}|}
\hline &  & \begin{minipage}[c][0.95cm][c]{2cm}\centering Source\\ tampering\end{minipage} & \multicolumn{1}{c|}{\begin{minipage}{2cm}\centering Binary\\ tampering\end{minipage}} & \multicolumn{1}{c|}{\begin{minipage}{2cm}\centering Improper build\end{minipage}} & \multicolumn{1}{c|}{\begin{minipage}{2cm}\centering  Control info\\tampering\end{minipage}} & \multicolumn{1}{c|}{\begin{minipage}{2cm}\centering Infrastructure tampering\end{minipage}} \\ \hline

\multicolumn{1}{|c|}{\multirow{4}{*}{\begin{minipage}{1.3cm}External Entities \end{minipage}}} & \fontDFD{Developer} & S, R & & & & \\ \cline{2-7} 
\multicolumn{1}{|c|}{} & \fontDFD{DevOps Eng} & & & & S, R & S, R \\ \cline{2-7} 
\multicolumn{1}{|c|}{} & \fontDFD{Tester} & & & & S, R & \\ \cline{2-7} 
\multicolumn{1}{|c|}{} & \fontDFD{User} & &  & & & \\ \hline

\multirow{6}{*}{\begin{minipage}{1.3cm}Processes\end{minipage}} 
& \fontDFD{Integration} & S, T, E & & & & \\ \cline{2-7} 
& \fontDFD{CI} & T & S, T, E & S, T, E & & \\ \cline{2-7} 
& \fontDFD{Test} & & & & S, T, E & \\ \cline{2-7} 
& \fontDFD{Deployment} & & S, T, E & & E & S, T, E \\ \cline{2-7} 
& \fontDFD{Release} & & S, T, E & & & \\ \cline{2-7} 
& \fontDFD{Download/ App Server} & & S, T, E & & & \\ \hline

\multirow{8}{*}{\begin{minipage}{1.3cm}Data Flows\tnote{1}\end{minipage}} 
& & \raggedright T: \sf Dev $\to$ I $\to$ VCS $\to$ CI & T: \sf CI $\to$ AR  & T: \sf PL $\to$ CI & T: \sf T $\to$ AR & \begin{minipage}[t]{2cm}\raggedright T: \sf DE $\to$ IR $\to$ D\end{minipage} \\ \cline{3-7} 
& & T: \sf PL $\to$ CI & \raggedright T: \sf AR $\to$ D $\to$ WA $\to$ R $\to$ BR $\to$ DS $\to$ U & & \raggedright T: \sf D $\to$ AR &  \\ \cline{3-7} 
& & & & & \raggedright T: \sf DE $\to$ IR $\to$ T & \\ \cline{3-7} 
& & & & & \raggedright T: \sf T $\to$ Tr & \\ \cline{3-7} 
& & & & & \raggedright T: \sf Tr $\to$ T & \\ \hline

\multirow{7}{*}{\begin{minipage}{1.3cm}Data Stores\end{minipage}} 
& \fontDFD{VCS} & T & & & & \\ \cline{2-7} 
& \fontDFD{Artifact Repo} & & T & & T & \\ \cline{2-7} 
& \fontDFD{Infrastruct Repo} & & & & T & T\\ \cline{2-7} 
& \fontDFD{Web/App Store} & & T & & & \\ \cline{2-7} 
& \fontDFD{Binary Repo} & & T & & & \\ \hline

\end{tabular}
\begin{tablenotes}
\item [1] Abbreviations used for data flows: \sf [AR]~Artifact Repository; [BR]~Binary Repository; [CI]~Continuous Integration; [D]~Deployment; [Dev]~Developer; [DE]~DevOps Engineer; [WA]~Directory on a Web server/App Store; [DS]~Download/App Server; [IR]~Infrastructure Repository; [I]~Integration; [PL]~Packages and Libraries; [R]~Release; [T]~Test; [Tr]~Tester; [U]~User; [VCS]~VCS. 
\end{tablenotes}
\end{threeparttable}
}

\end{table}

%% file: chapters/5_CaseStudy.tex
\section{Case Study: Applying the Threat Model to a Deployment Pipeline} \label{cap:modelagemBass}

One of our objectives is to perform the analysis of at least one publicly documented pipeline. Therefore, to show the applicability of the model,  we chose a pipeline that has already been published in the literature.

Thus, this section presents threat modeling for the deployment pipeline presented by \cite{bass2015securing}. First, Section~\ref{sub:pipelineOverview} describes the pipeline, and Section~\ref{sub:pipelineModeling} presents its DFD model. Section~\ref{sub:pipelineThreatModel} shows how the threat model from Section~\ref{cap:modelagemAmeaca} is applied to this pipeline.
Finally, Section~\ref{sub:pipelineDiscussion} discusses this case study.

\subsection{Deployment pipeline overview} \label{sub:pipelineOverview}

\begin{figure}
    \centering
    \caption{Deployment pipeline. Adapted from \cite{bass2015securing}}
    \includegraphics[width=1.1\linewidth]{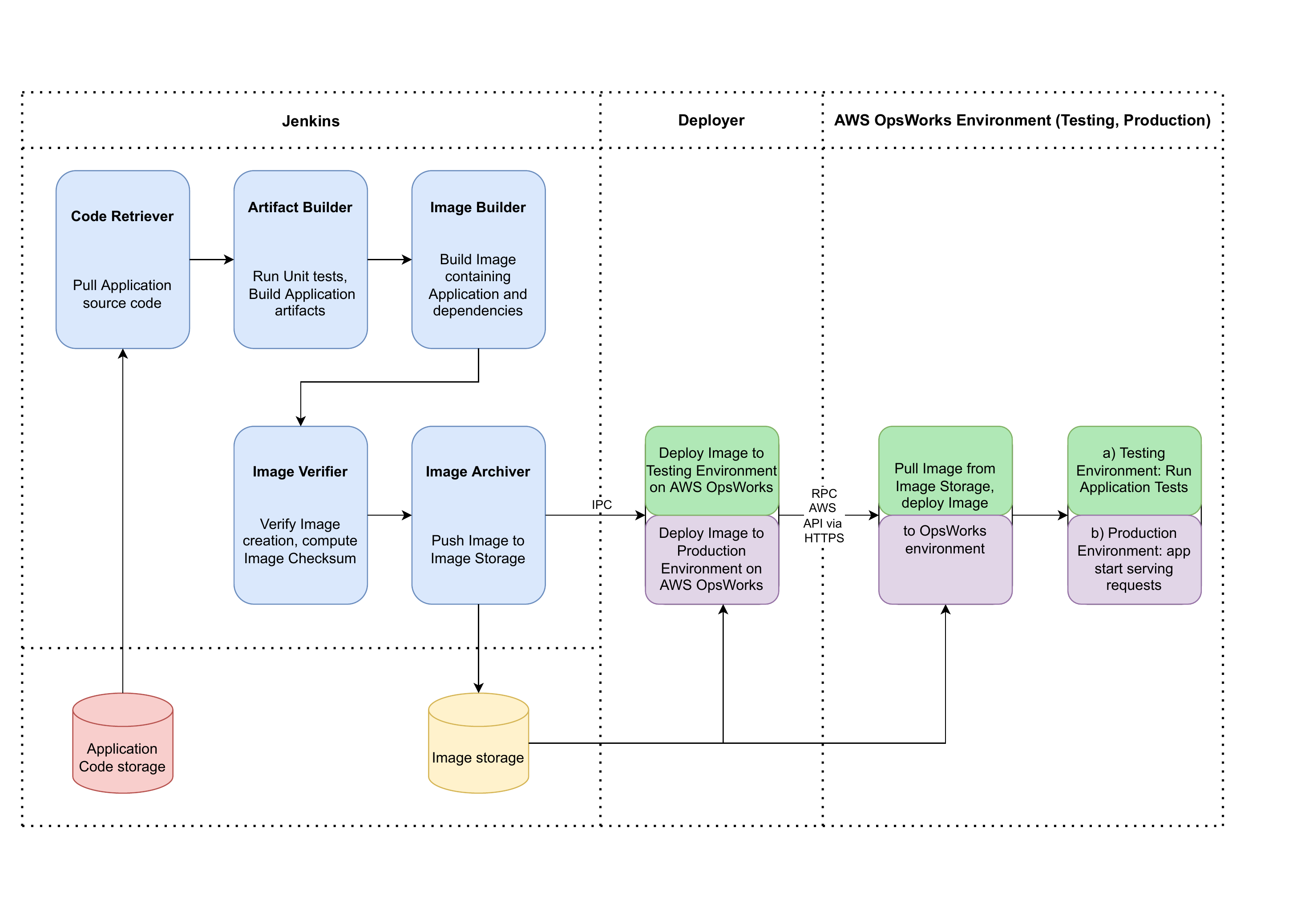}
    \label{fig:Pipeline_Bass}
\end{figure}

Figure \ref{fig:Pipeline_Bass} shows the deployment pipeline presented by \cite{bass2015securing}; we added colors to the elements to aid in mapping this pipeline to the DFD from Section~\ref{sec:modelagemPipeline}. This pipeline involves the Continuous Integration, Infrastructure as Code, Deployment, and Release stages. It is based on Jenkins and deploys into AWS\@. However, the steps are generic and must be performed by any Continuous Integration/Deployment tool suite.

Jenkins is an open-source automation server that supports building, deploying, and automating projects \cite{jenkins2016}. Jenkins can be used to perform several steps in a continuous deployment pipeline. These steps can include pulling source code from a repository, building the application binary from source code, running tests, creating an image, and storing the image in a repository \cite{rimba2015composing}. In this pipeline, the stores are implemented as AWS Simple Storage Service (S3) buckets.

In the pipeline shown in Figure~\ref{fig:Pipeline_Bass}, we can see that Jenkins represents the Continuous Integration step and has five components:

\begin{itemize}
    \item \textbf{Code Retriever}: responsible for pulling the application's source code, which is stored in a repository (Application Code storage). This repository can be a version control system;
    \item \textbf{Artifact Builder}: in charge of running unit tests and building the application's artifacts;
    \item \textbf{Image Builder}: responsible for packaging application binaries into an image;
    \item  \textbf{Image Verifier}: responsible for verifying the created image, and computing the image checksum; and
    \item \textbf{Image Archiver}: in charge of pushing the image to a repository (Image Storage) and calling a Deployer.
\end{itemize}

Deployer and AWS OpsWorks Environment represent the Infrastructure as Code, Deployment and Release steps. Deployer is critical for setting up test environments. This setup includes installing the application and its dependencies inside a virtual machine, configuring and running the application, and initiating the tests \cite{rimba2015composing}. AWS OpsWorks is responsible for helping with environment setup and application installation.

There are several types of tests to ensure that the application has the desired functionality. These tests include unit testing, integration testing, and end-to-end testing. If the image passes all the automated tests, a Deployer can deploy it to the production environment and make it available to users.

\subsection{Modeling the deployment pipeline} \label{sub:pipelineModeling}


To model the deployment pipeline presented in Section~\ref{sub:pipelineOverview}, we adapted the DFD from Section~\ref{sec:modelagemPipeline} to match the characteristics of this pipeline, instead of developing a model from scratch. This is key to making the modeling process simpler and faster, and results in a DFD that makes it easier to adapt the threat model to a new pipeline.

\begin{figure}
    \centering
    \caption{Color mapping between the pipeline elements of Figure~\ref{fig:Pipeline_Bass} and the DFD shown in Figure~\ref{fig:DFD_Pipeline} (page~\pageref{fig:DFD_Pipeline})}
    \includegraphics[width=1.4\linewidth, angle = 90]{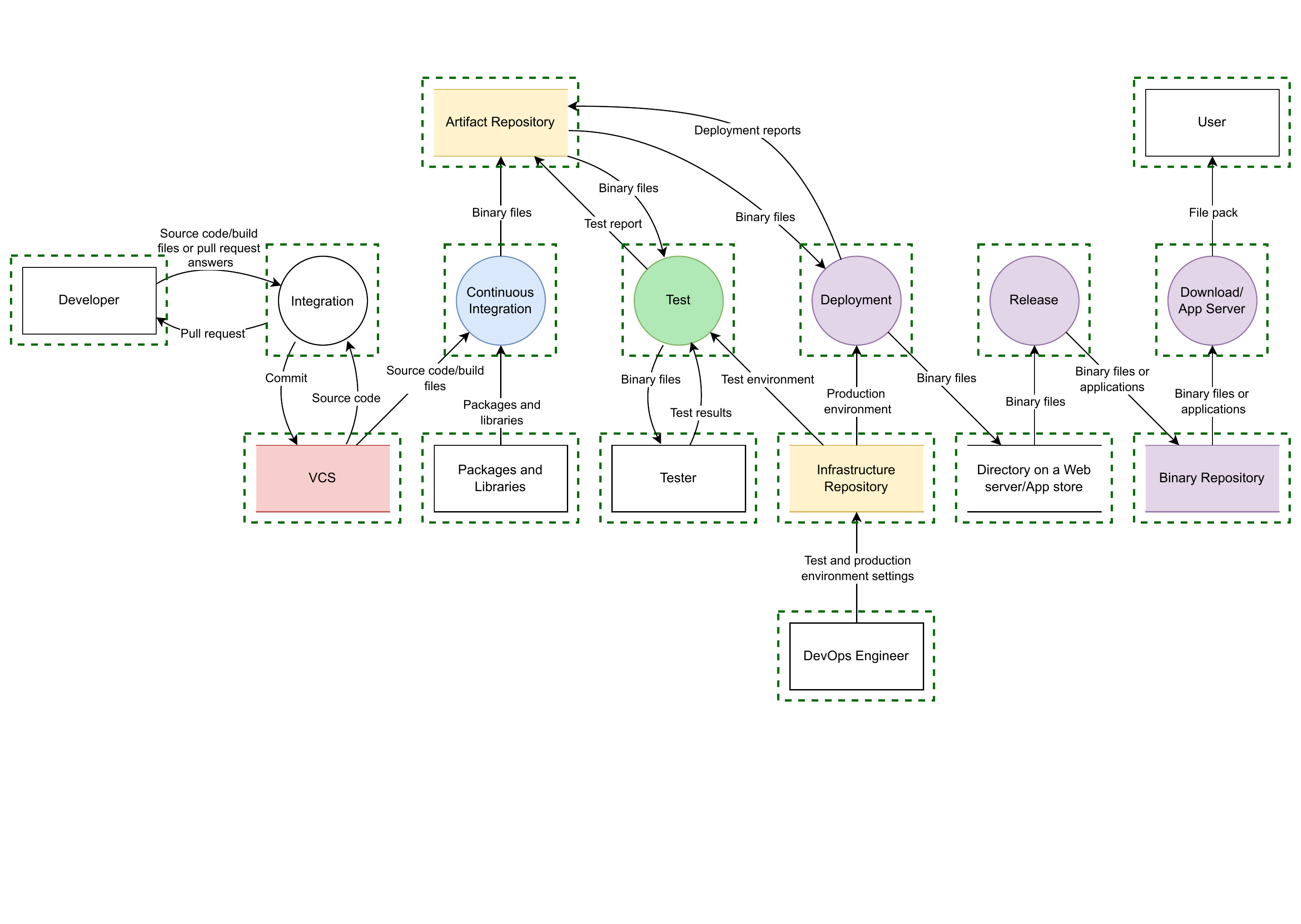}
    \label{fig:map_dfd}
\end{figure}

First, we perform a color mapping between the pipeline elements of Section~\ref{sub:pipelineOverview} and the DFD shown in Section~\ref{sec:modelagemPipeline}. Figure~\ref{fig:map_dfd} presents the result of this mapping. We can see that the Jenkins elements (Code Retriever, Artifact Builder, Image Builder, Image Verifier, and Image Archiver) represent the \fontDFD{Continuous Integration} step. The Application Code storage repository is equivalent to the \fontDFD{VCS} data store. Image Storage represents AWS S3 buckets. Therefore, it is responsible for storing the image, settings and logs of the test and production environment. Thus, it is equivalent to the \fontDFD{Artifact} and \fontDFD{Infrastructure Repository}. Finally, Deployer and AWS OpsWorks represent the \fontDFD{Test}, \fontDFD{Deployment}, and \fontDFD{Release} processes. Inside AWS we have a virtual machine (VM), which has some local storage that corresponds to the \fontDFD{Binary Repository}.

\begin{figure}[!htb]
    \centering
    \caption{DFD for the deployment pipeline}
    \includegraphics[width=1.2\linewidth]{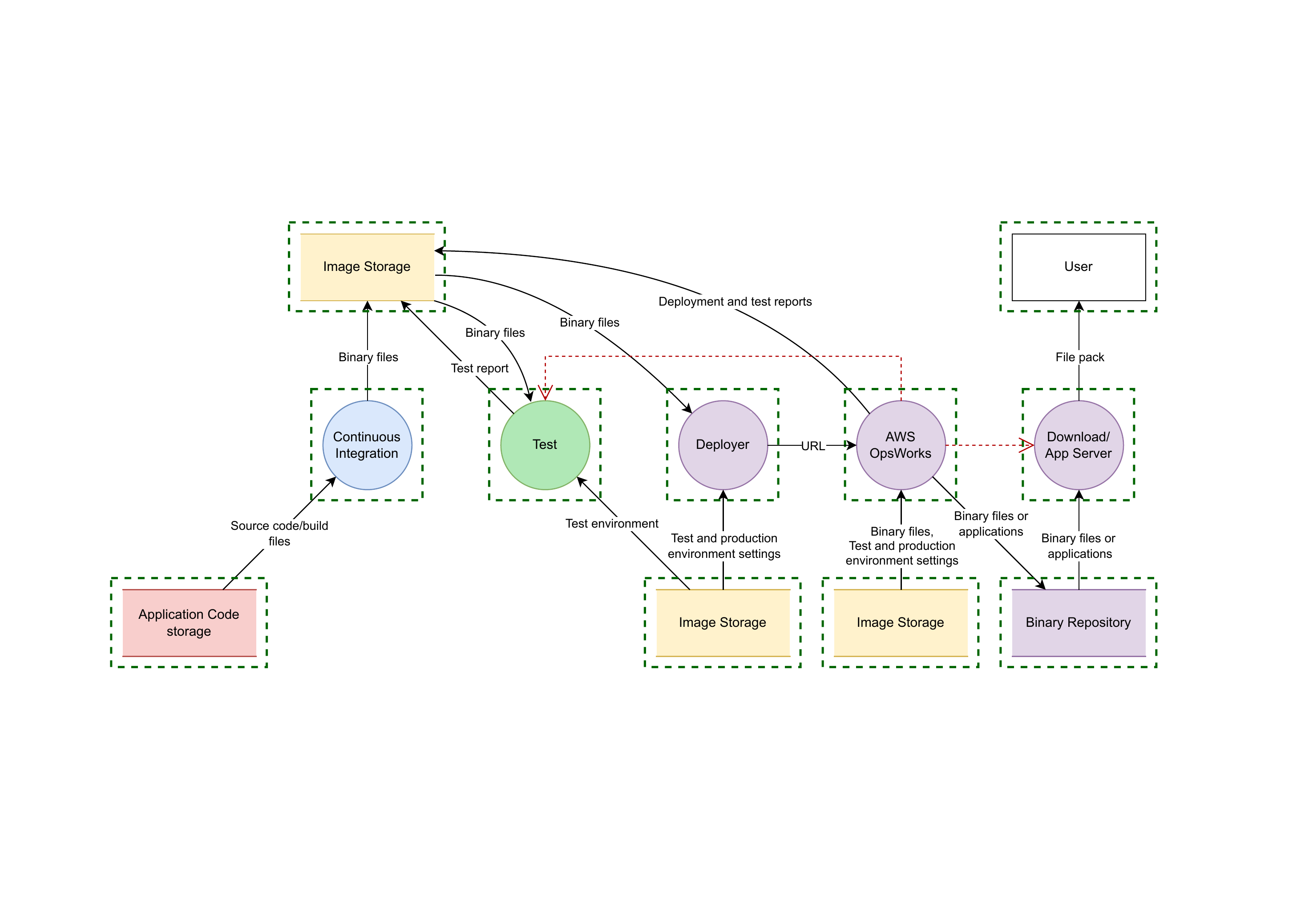}
    \label{fig:map_final_dfd}
\end{figure}

Then we remove elements from the diagram that are considered out of scope in \cite{bass2015securing} and adapt the data flows according to the deployment pipeline. The result of these modifications can be seen in Figure~\ref{fig:map_final_dfd}. The \fontDFD{Continuous Integration} process pulls the source code from the \fontDFD{Application Code storage} repository, builds the binary code (going through all the steps colored in blue in Figure~\ref{fig:Pipeline_Bass}), stores the image in the \fontDFD{Image Storage} data store, and triggers a \fontDFD{Deployer}.

Once triggered, \fontDFD{Deployer} generates a temporary URL of the location of the image in a bucket. This URL is encrypted. Then the \fontDFD{Deployer} invokes \fontDFD{AWS OpsWorks}, which can pull the image from the bucket specified in the URL, and run it on the desired instance. \fontDFD{AWS OpsWorks} instantiates the \fontDFD{Test} and \fontDFD{Download/App Server} processes. Since control flows are not usually  represented in data flow diagrams, we depict these flows as red dashed arrows. After the automated tests are finished and the image is deployed, the respective log reports are stored in \fontDFD{Image Storage}. Finally, the \fontDFD{Download/App Server} is the server from which the \fontDFD{User} can access the software.

The trust boundaries (delimited by dashed lines) are determined according to the context. \fontDFD{AWS OpsWorks}, \fontDFD{Binary Repository}, \fontDFD{Download/App Server}, and \fontDFD{Test} are considered within the same trust boundary because they are considered a single service. On the other hand, \fontDFD{Application Code storage}, \fontDFD{Continuous Integration}, \fontDFD{Deployer}, and \fontDFD{Image Storage} have their own trust boundaries as they present different contexts from the other elements.

\subsection{Threat Model for the Deployment Pipeline} \label{sub:pipelineThreatModel}

To develop a threat model for the deployment pipeline described in Section~\ref{sub:pipelineOverview}, we mapped the threats presented in Section~\ref{sec:modelagemAmeacas} to the DFD introduced in Section~\ref{sub:pipelineModeling}. So, we show the results of this threat mapping in Sections~\ref{subsub:processes_pipeline_Bass} to \ref{subsub:stores_pipeline_Bass}.

The original threat model presented in \cite{bass2015securing} excludes some threats, including:
\begin{itemize}
\item elevation of privilege threats;
\item threats from third-party packages and libraries;
\item malicious build systems; and
\item malicious cloud infrastructures.
\end{itemize}
Therefore, to arrive at a threat model that can be directly compared to the original one, we chose to disregard such threats as well; after all, it would be unfair to say that the original model overlooks threats that have been deliberately excluded. However, we note that the threats introduced in Section~\ref{sec:modelagemAmeacas} that fall into these excluded categories still apply to this pipeline.

\subsubsection{Processes} \label{subsub:processes_pipeline_Bass}

The DFD has five processes: \fontDFD{Continuous Integration}, \fontDFD{Test}, \fontDFD{Deployer}, \fontDFD{AWS OpsWorks}, and \fontDFD{Download/App Server}. These processes are subject to some common threats that can be mitigated in the same way, albeit with different consequences for each process. The threats found are:

\begin{itemize}
    \item \textit{Spoofing}: server spoofing -- the possibility of an illegitimate process, data store, or external entity posing as the real one and supplying malicious data. The possible consequences are:
    
    \begin{itemize}
        \item source tampering:
        \begin{itemize}
            \item via spoofed \fontDFD{Application Code storage} data store (against the \fontDFD{Continuous Integration} process);
            \item via spoofed \fontDFD{Continuous Integration} (Code Retriever against Artifact Builder, Artifact Builder against Image Builder);
        \end{itemize}
        \item binary tampering:
        \begin{itemize}
            \item via spoofed \fontDFD{Deployer} (against \fontDFD{AWS OpsWorks});
            \item via spoofed \fontDFD{Download/App Server} (against \fontDFD{User});
             \item via spoofed \fontDFD{Continuous Integration} (Image Builder against Image Archiver, Image Archiver against \fontDFD{Image Storage});
            
            \item via spoofed \fontDFD{Image Storage} (against \fontDFD{Deployer} and \fontDFD{AWS OpsWorks});
        \end{itemize}
        \item control information tampering:
        \begin{itemize}
            \item via spoofed \fontDFD{Test} (against \fontDFD{Image Storage});
            \item via spoofed \fontDFD{Deployer} (against \fontDFD{AWS OpsWorks});
            \item via spoofed \fontDFD{AWS OpsWorks} (against \fontDFD{Image Storage});
            \item via spoofed \fontDFD{Image Storage} (against \fontDFD{Test}, \fontDFD{Deployer}, and \fontDFD{AWS OpsWorks});
        \end{itemize}
        \item infrastructure tampering:
        \begin{itemize}
            \item via spoofed \fontDFD{Deployer} (against \fontDFD{AWS OpsWorks});
            \item via spoofed \fontDFD{Image Storage} (against \fontDFD{Deployer} and \fontDFD{AWS OpsWorks}).
        \end{itemize}
    \end{itemize}

    Threats that cause control information tampering, which are related to test and deployment reports, are not considered in \cite{bass2015securing}. Therefore, we noted additional threats to the model. The threat via spoofed \fontDFD{Download/App Server} is particularly dangerous: users are often unable to verify that the application they are interacting with is trustworthy \citeonline{thelinuxfoundation2020}.

    \item \textit{Tampering}: the processes do not know whether the data they are receiving, whether from external entities, data stores, or other processes, is trustworthy. This threat affects all processes, with different consequences depending on the origin of the data and the receiving process. For example, malicious input to \fontDFD{Continuous Integration} may result in source tampering (Code Retriever, Artifact Builder) or binary tampering (Image Builder, Image Archiver).

    Another tampering threat that is present is local falsification. This may lead to any of the consequences defined in Section~\ref{sec:threat-consequences} (except improper build, since \cite{bass2015securing} assumes a correct build system), depending on the process. For instance, a malicious \fontDFD{Continuous Integration} (Image Builder, Image Archiver) process may perform binary tampering, while a malicious \fontDFD{Deployer} process may cause binary tampering, control information tampering, and/or infrastructure tampering.

\end{itemize}

Section~\ref{sub:processos} also introduces threats to the \fontDFD{Continuous Integration} process that relate to third-party packages and libraries. Since packages and libraries are deemed out of scope in \cite{bass2015securing}, we do not consider them here.

\subsubsection{Data flows} \label{subsub:flows_pipeline_Bass}

The data flows in the diagram are susceptible to common threats, the consequences of which differ slightly depending on the source and destination of each flow. However, the threats do not apply to flows within the same trust boundary, namely the flows between \fontDFD{AWS OpsWorks}, \fontDFD{Binary Repository}, and \fontDFD{Download/App Server}. The threats found are:

\begin{itemize}
    \item \textit{Tampering}: an attacker may alter data during communication, with the following possible consequences:
    \begin{itemize}
        \item source tampering:
        \begin{itemize}
            \item \fontDFD{Application Code storage} $\to$ \fontDFD{Continuous Integration} (Code Retriever);
        \end{itemize}
        
        \item binary tampering: any flow whose tampering enables an attacker to alter binaries, installation images, or tools that govern the image building process:
        \begin{itemize}
            \item \fontDFD{Continuous Integration} (Image Archiver) $\to$ \fontDFD{Image Storage};
            \item \fontDFD{Image Storage} $\to$ \fontDFD{Deployer} $\to$ \fontDFD{AWS OpsWorks}; 
            \item \fontDFD{Image Storage} $\to$ \fontDFD{AWS OpsWorks}; 
        \end{itemize}
       
        \item control information tampering:
        \begin{itemize}
            \item \fontDFD{Test} $\to$ \fontDFD{Image Storage};
            \item \fontDFD{AWS OpsWorks} $\to$ \fontDFD{Image Storage};
            \item \fontDFD{Image Storage} $\to$ \fontDFD{Deployer};
            \item \fontDFD{Image Storage} $\to$ \fontDFD{AWS OpsWorks};
            \item \fontDFD{Image Storage} $\to$ \fontDFD{Test}.
        \end{itemize}
        \item infrastructure tampering:
        \begin{itemize}
            \item \fontDFD{Image Storage} $\to$ \fontDFD{Deployer};
            \item \fontDFD{Image Storage} $\to$ \fontDFD{AWS OpsWorks}.
        \end{itemize}
    \end{itemize}

\end{itemize}

\subsubsection{Data stores} \label{subsub:stores_pipeline_Bass}

There are three data stores in the DFD: \fontDFD{Application Code storage}, \fontDFD{Image Storage}, and \fontDFD{Binary Repository}. For each one similar threats were found, albeit with different consequences. However, the threats do not apply to the \fontDFD{Binary Repository} data store because \cite{bass2015securing} assumes the AWS infrastructure is trustworthy. The threats found are:

\begin{itemize}
     \item \textit{Tampering}: an attacker may tamper with stored data. The \fontDFD{Application Code storage} repository stores source code, and its compromise can lead to source tampering. In the \fontDFD{Image Storage}, changes in test reports can make it possible to legitimate a tampered software (control information tampering), or a correct image can be replaced by a malicious one (binary tampering). The \fontDFD{Image Storage} also stores configuration files for test and deployment infrastructures, and changes to these files may be used to legitimate tampered software (control information tampering) or introduce vulnerabilities in the deployment infrastructure (infrastructure tampering). 
     
     As mentioned in Section~\ref{subsub:processes_pipeline_Bass}, threats related to test/deployment reports and configuration data are not considered in \cite{bass2015securing}.

\end{itemize}

\subsection{Discussion} \label{sub:pipelineDiscussion}

To demonstrate the usefulness of the threat model introduced in Section~\ref{cap:modelagemAmeaca}, we showed how it could be applied to the deployment pipeline described in \cite{bass2015securing}. Rather than developing a threat model from scratch, we mapped our model to this pipeline, identifying corresponding elements and removing threats that are no longer relevant.

Table~\ref{tab:summary_model_bass} summarizes the threat model, and Table~\ref{tab:consequencias_Bass} provides a complementary view of the threat consequences for each DFD element. There are fewer threats in this model (five) than in our generic model (13) since some threats had been excluded from consideration in the original threat model in \cite{bass2015securing}, while others were dismissed for irrelevance. The components that correspond to the \fontDFD{Continuous Integration} process in our model -- Artifact Builder, Image Builder, Image Verifier, and Image Builder -- face fewer threat consequences, partly because improper builds are deemed out of scope and partly because partitioning functionality across multiple components reduces the individual attack surfaces. The \fontDFD{Deployer} and \fontDFD{AWS OpsWorks} processes are subject to a significant number of threat consequences, just like their counterparts in the generic model, \fontDFD{Deployment} and \fontDFD{Release}.

We highlight that by applying our model within the scope adopted in \cite{bass2015securing}, we found the same threats originally identified for this pipeline, while also uncovering additional ones related to logs and configuration data from test and production environments. We also note the many threats that may lead to control information tampering, and which were overlooked in \cite{bass2015securing}.

Thus, instantiating our generic model for this specific pipeline produced a useful threat model with limited effort. However, we note that this threat model considers the pipeline at the design level, and does not take into consideration implementation vulnerabilities.

\input{chapters/tab_threats_deploy_pipeline}

\begin{table}[h!]
\centering
\caption{Summary of threat consequences found for the deployment pipeline}
\label{tab:consequencias_Bass}
\def\arraystretch{1.3}
\scalebox{0.85}{
\begin{threeparttable}[b]
\begin{tabular}{|p{1.3cm}|p{2.6cm}|p{2cm}|p{2cm}|p{2cm}|p{2cm}|}
\hline &  & \multicolumn{1}{c|}{\begin{minipage}{2cm}Source\\tampering\end{minipage}} & \multicolumn{1}{c|}{\begin{minipage}{2cm}Binary\\tampering\end{minipage}} & \multicolumn{1}{c|}{\begin{minipage}{2cm}Control info tampering\end{minipage}} & \multicolumn{1}{c|}{\begin{minipage}{2cm}Infrastructure tampering\end{minipage}} \\ \hline

\multirow{9}{*}{\begin{minipage}{1.3cm}Processes\end{minipage}} 
& \fontDFD{Code Retriever}   & S, T &  &  & \\ \cline{2-6} 
& \fontDFD{Artifact Builder} & S, T & T &  & \\ \cline{2-6} 
& \fontDFD{Image Builder}    &  & S, T &  & \\ \cline{2-6} 
& \fontDFD{Image Verifier}   &  &  &  & \\ \cline{2-6}
& \fontDFD{Image Archiver}   &  & S, T &  & \\ \cline{2-6} 
& \fontDFD{Test}             & & &  S, T & \\ \cline{2-6} 
& \fontDFD{Deployer}         & & S, T &  S, T & S, T \\ \cline{2-6} 
& \fontDFD{AWS OpsWorks}     & & T &  S, T & T \\ \cline{2-6} 
& \fontDFD{Download/App Server} & & S, T  & & \\ \hline

\multirow{5}{*}{\begin{minipage}{1.3cm}Data Flows\tnote{1} \end{minipage}} 
& & T: ACS $\to$ CR  & T: IA $\to$ IS  &   T: T $\to$ IS & T: IS $\to$ D \\ \cline{3-6} 
& & & T: IS $\to$ D $\to$ AWS & T: IS $\to$ T & T: IS $\to$ AWS \\ \cline{3-6} 
& & &  T: IS $\to$ AWS &  T: AWS $\to$ IS &   \\ \cline{3-6} 
& & &  &  T: IS $\to$ D & \\ \cline{3-6} 
& & & &  T: IS $\to$ AWS & \\ \hline

\multirow{3}{*}{\begin{minipage}{1.3cm}Data Stores\end{minipage}} 
& \fontDFD{App Code Storage}  & T & &  & \\ \cline{2-6} 
& \fontDFD{Image Storage}     & & T &  T & T \\ \cline{2-6} 
& \fontDFD{Binary Repository} & &  & & \\ \hline

\end{tabular}
\begin{tablenotes}
\item [1] Abbreviations used for data flows: \sf [ACS]~Application Code Storage; [AB]~Artifact Builder;  [AWS]~AWS OpsWorks; [BR]~Binary Repository; [CR]~Code Retriever; [D]~Deployer; [AS]~Download/App Server; [IA]~Image Archiver; [IB]~Image Builder; [IS]~Image Storage; [IV]~Image Verifier; [T]~Test.
\end{tablenotes}
\end{threeparttable}
}
\end{table}

%% file: chapters/tab_threats_deploy_pipeline.tex
\begin{table}[h!]
\centering
\caption{Summary of threats found for the deployment pipeline}
\label{tab:summary_model_bass}
\def\arraystretch{1.3}
\scalebox{1}{
    \begin{tabular}{|p{5cm}|p{3cm}|p{4.5cm}|}
    \hline
    \multicolumn{1}{|c|}{\textbf{Diagram element}} & \multicolumn{1}{c|}{\textbf{Threat type}} & \multicolumn{1}{c|}{\textbf{Threat}} \\ \hline
    
    \multirow{3}{*}{\begin{minipage}{5cm} \fontDFD{Continuous Integration}\\\fontDFD{Test}\\\fontDFD{Deployer}\\\fontDFD{AWS OpsWorks}\\\fontDFD{Download/App Server}\end{minipage}} & \begin{minipage}[t][0.6cm][t]{3cm} Spoofing \end{minipage} & Server spoofing  \\ \cline{2-3} 
    & \multirow{2}{*}{Tampering} & \begin{minipage}[t][0.6cm][t]{4.5cm}Receiving unreliable data\end{minipage} \\ \cline{3-3} &   & \begin{minipage}[t][0.6cm][t]{4.5cm} Local spoofing \end{minipage} \\ \hline
    
    Data flows (except \fontDFD{AWS OpsWorks} $\to$ \fontDFD{Binary Repository} $\to$ \fontDFD{Download/App Server}) & Tampering & Changing data during communication \\ \hline
    
    \begin{minipage}{5cm}\fontDFD{Application Code storage}\\\fontDFD{Image Storage}\end{minipage}  & \begin{minipage}[t][0.8cm][t]{3cm} Tampering \end{minipage} & Improper data alteration \\ \hline
    
    \end{tabular}
}
\end{table}

%% file: chapters/6_Conclusion.tex
\section{Conclusion} \label{cap:conclusao}

In recent years, we have seen an increase in the introduction of vulnerabilities via the compromise of software development pipelines -- the infrastructures used to build, deploy, and deliver software --, giving rise to supply chain attacks. Improving software supply chain security demands secure development pipelines, and, to build a secure pipeline, we need a comprehensive view of the relevant threats. In this paper, we review documented attacks and introduce a systematic threat model for generic software development pipelines. The model is focused on integrity threats and was developed using the STRIDE framework. Our model is complemented by a discussion of suitable mitigations for the threats. We also present a case study describing how our model can be applied to a specific pipeline with limited effort, producing favorable results.

The research community can leverage our work to find avenues for improving the security of software development pipelines and supply chains. We note that validating that software components have not been compromised remains an elusive goal; this is especially true for hosted applications as for installable software several mitigations have already been proposed. Infrastructure-as-code security has received little research attention, and further work is needed. Our results are also directly relevant to the software industry: software development organizations can use our threat model and proposed mitigations to enhance the security of their development pipelines.  

In future work, we aim to apply our threat model to other software development pipelines. We also intend to investigate the security of software supply chains organized as interconnected development pipelines.

%% file: main.bbl
\begin{thebibliography}{117}
\expandafter\ifx\csname natexlab\endcsname\relax\def\natexlab#1{#1}\fi
\providecommand{\url}[1]{\texttt{#1}}
\providecommand{\href}[2]{#2}
\providecommand{\path}[1]{#1}
\providecommand{\DOIprefix}{doi:}
\providecommand{\ArXivprefix}{arXiv:}
\providecommand{\URLprefix}{URL: }
\providecommand{\Pubmedprefix}{pmid:}
\providecommand{\doi}[1]{\href{http://dx.doi.org/#1}{\path{#1}}}
\providecommand{\Pubmed}[1]{\href{pmid:#1}{\path{#1}}}
\providecommand{\bibinfo}[2]{#2}
\ifx\xfnm\relax \def\xfnm[#1]{\unskip,\space#1}\fi
\bibitem[{Abomhara et~al.(2015)Abomhara, Gerdes and
  K{\o}ien}]{abomhara2015:stride}
\bibinfo{author}{Abomhara, M.}, \bibinfo{author}{Gerdes, M.},
  \bibinfo{author}{K{\o}ien, G.M.}, \bibinfo{year}{2015}.
\newblock \bibinfo{title}{A {STRIDE}-based threat model for telehealth
  systems}.
\newblock \bibinfo{journal}{Norsk informasjonssikkerhetskonferanse (NISK)}
  \bibinfo{volume}{8}, \bibinfo{pages}{82--96}.
\bibitem[{Adams and McIntosh(2016)}]{adams2016modern}
\bibinfo{author}{Adams, B.}, \bibinfo{author}{McIntosh, S.},
  \bibinfo{year}{2016}.
\newblock \bibinfo{title}{Modern release engineering in a nutshell -- why
  researchers should care}, in: \bibinfo{booktitle}{2016 IEEE 23rd
  International Conference on Software Analysis, Evolution, and Reengineering
  (SANER)}, \bibinfo{organization}{IEEE}. pp. \bibinfo{pages}{78--90}.
\newblock \DOIprefix\doi{10.1109/SANER.2016.108}.
\bibitem[{Al~Sabbagh and Kowalski(2015)}]{al2015socio}
\bibinfo{author}{Al~Sabbagh, B.}, \bibinfo{author}{Kowalski, S.},
  \bibinfo{year}{2015}.
\newblock \bibinfo{title}{A socio-technical framework for threat modeling a
  software supply chain}.
\newblock \bibinfo{journal}{IEEE Security \& Privacy} \bibinfo{volume}{13},
  \bibinfo{pages}{30--39}.
\bibitem[{Allen et~al.(2008)Allen, Barnum, Ellison, McGraw and
  Mead}]{allen2008software}
\bibinfo{author}{Allen, J.H.}, \bibinfo{author}{Barnum, S.},
  \bibinfo{author}{Ellison, R.J.}, \bibinfo{author}{McGraw, G.},
  \bibinfo{author}{Mead, N.R.}, \bibinfo{year}{2008}.
\newblock \bibinfo{title}{Software security engineering}.
\newblock \bibinfo{publisher}{Addison-Wesley}.
\bibitem[{Barabanov et~al.(2020)Barabanov, Markov and
  Tsirlov}]{barabanov2020systematics}
\bibinfo{author}{Barabanov, A.}, \bibinfo{author}{Markov, A.},
  \bibinfo{author}{Tsirlov, V.}, \bibinfo{year}{2020}.
\newblock \bibinfo{title}{On systematics of the information security of
  software supply chains}, in: \bibinfo{booktitle}{Proceedings of the
  Computational Methods in Systems and Software},
  \bibinfo{organization}{Springer}. pp. \bibinfo{pages}{115--129}.
\bibitem[{Barabanov et~al.(2018)Barabanov, Markov, Grishin and
  Tsirlov}]{barabanov2018current}
\bibinfo{author}{Barabanov, A.V.}, \bibinfo{author}{Markov, A.S.},
  \bibinfo{author}{Grishin, M.I.}, \bibinfo{author}{Tsirlov, V.L.},
  \bibinfo{year}{2018}.
\newblock \bibinfo{title}{Current taxonomy of information security threats in
  software development life cycle}, in: \bibinfo{booktitle}{2018 IEEE 12th
  International Conference on Application of Information and Communication
  Technologies (AICT)}, pp. \bibinfo{pages}{1--6}.
\newblock \DOIprefix\doi{10.1109/ICAICT.2018.8747065}.
\bibitem[{Bass et~al.(2015)Bass, Holz, Rimba, Tran and Zhu}]{bass2015securing}
\bibinfo{author}{Bass, L.}, \bibinfo{author}{Holz, R.}, \bibinfo{author}{Rimba,
  P.}, \bibinfo{author}{Tran, A.B.}, \bibinfo{author}{Zhu, L.},
  \bibinfo{year}{2015}.
\newblock \bibinfo{title}{Securing a deployment pipeline}, in:
  \bibinfo{booktitle}{2015 IEEE/ACM 3rd International Workshop on Release
  Engineering}, \bibinfo{organization}{IEEE}. pp. \bibinfo{pages}{4--7}.
\newblock \DOIprefix\doi{10.1109/RELENG.2015.11}.
\bibitem[{Birsan(2021)}]{birsan2021dependency}
\bibinfo{author}{Birsan, A.}, \bibinfo{year}{2021}.
\newblock \bibinfo{title}{Dependency confusion: How {I} hacked into {Apple},
  {Microsoft} and {Dozens} of other companies}.
\newblock \bibinfo{howpublished}{Medium}.
\newblock \URLprefix
  \url{https://medium.com/@alex.birsan/dependency-confusion-4a5d60fec610}.
\bibitem[{Bitbucket(2021)}]{bit2021pull}
\bibinfo{author}{Bitbucket}, \bibinfo{year}{2021}.
\newblock \bibinfo{title}{Making a pull request}.
\newblock \bibinfo{howpublished}{Atlassian Bitbucket}.
\newblock \URLprefix
  \url{https://www.atlassian.com/git/tutorials/making-a-pull-request}.
\bibitem[{Brumaghin et~al.(2017)Brumaghin, Gibb, Mercer, Molyett and
  Williams}]{brumaghin2017}
\bibinfo{author}{Brumaghin, E.}, \bibinfo{author}{Gibb, R.},
  \bibinfo{author}{Mercer, W.}, \bibinfo{author}{Molyett, M.},
  \bibinfo{author}{Williams, C.}, \bibinfo{year}{2017}.
\newblock \bibinfo{title}{{CCleanup}: A vast number of machines at risk}.
\newblock \bibinfo{howpublished}{Talos Blog}.
\newblock \URLprefix
  \url{https://blog.talosintelligence.com/2017/09/avast-distributes-malware.html}.
\bibitem[{BSIMM(2022)}]{BSIMM2022}
\bibinfo{author}{BSIMM}, \bibinfo{year}{2022}.
\newblock \bibinfo{title}{{BSIMM} foundations report, version 13}.
\newblock \URLprefix
  \url{https://www.bsimm.com/content/dam/bsimm/reports/bsimm13-foundations.pdf}.
\bibitem[{Cagnazzo et~al.(2018)Cagnazzo, Hertlein, Holz and
  Pohlmann}]{Cagnazzo2018:threatmod}
\bibinfo{author}{Cagnazzo, M.}, \bibinfo{author}{Hertlein, M.},
  \bibinfo{author}{Holz, T.}, \bibinfo{author}{Pohlmann, N.},
  \bibinfo{year}{2018}.
\newblock \bibinfo{title}{Threat modeling for mobile health systems}, in:
  \bibinfo{booktitle}{2018 IEEE Wireless Communications and Networking
  Conference Workshops (WCNCW)}, pp. \bibinfo{pages}{314--319}.
\newblock \DOIprefix\doi{10.1109/WCNCW.2018.8369033}.
\bibitem[{Cimpanu(2017)}]{catalincimpanu2017}
\bibinfo{author}{Cimpanu, C.}, \bibinfo{year}{2017}.
\newblock \bibinfo{title}{{JavaScript} packages caught stealing environment
  variables}.
\newblock \bibinfo{howpublished}{Bleeping Computer}.
\newblock \URLprefix
  \url{https://www.bleepingcomputer.com/news/security/javascript-packages-caught-stealing-environment-variables/}.
\bibitem[{Clark and van Oorschot(2013)}]{clark2013sok}
\bibinfo{author}{Clark, J.}, \bibinfo{author}{van Oorschot, P.C.},
  \bibinfo{year}{2013}.
\newblock \bibinfo{title}{{SoK: SSL and HTTPS: Revisiting Past Challenges and
  Evaluating Certificate Trust Model Enhancements}}, in:
  \bibinfo{booktitle}{2013 IEEE Symposium on Security and Privacy},
  \bibinfo{organization}{IEEE}. pp. \bibinfo{pages}{511--525}.
\newblock \DOIprefix\doi{10.1109/SP.2013.41}.
\bibitem[{Corbet(2003)}]{corbet2003:backdoor}
\bibinfo{author}{Corbet, J.}, \bibinfo{year}{2003}.
\newblock \bibinfo{title}{An attempt to backdoor the kernel}.
\newblock \URLprefix \url{https://lwn.net/Articles/57135/}.
\bibitem[{Corfield(2022)}]{garethcorfield2022phishing}
\bibinfo{author}{Corfield, G.}, \bibinfo{year}{2022}.
\newblock \bibinfo{title}{Phishing kits' use of man-in-the-middle reverse
  proxies is growing, warns proofpoint}.
\newblock \bibinfo{howpublished}{The Register}.
\newblock \URLprefix
  \url{https://www.theregister.com/2022/02/03/proofpoint_mitm_reverse_proxies/}.
\bibitem[{Costa(2019)}]{costa2019}
\bibinfo{author}{Costa, T.}, \bibinfo{year}{2019}.
\newblock \bibinfo{title}{strong\_password v0.0.7 rubygem hijacked}.
\newblock \URLprefix
  \url{https://withatwist.dev/strong-password-rubygem-hijacked.html}.
\bibitem[{Diaz-Sanchez et~al.(2019)Diaz-Sanchez, Marin-Lopez, Mendoza, Cabarcos
  and Sherratt}]{diaz2019tls}
\bibinfo{author}{Diaz-Sanchez, D.}, \bibinfo{author}{Marin-Lopez, A.},
  \bibinfo{author}{Mendoza, F.A.}, \bibinfo{author}{Cabarcos, P.A.},
  \bibinfo{author}{Sherratt, R.S.}, \bibinfo{year}{2019}.
\newblock \bibinfo{title}{{TLS/PKI Challenges and Certificate Pinning
  Techniques for IoT and M2M Secure Communications}}.
\newblock \bibinfo{journal}{IEEE Communications Surveys \& Tutorials}
  \bibinfo{volume}{21}, \bibinfo{pages}{3502--3531}.
\newblock \DOIprefix\doi{10.1109/COMST.2019.2914453}.
\bibitem[{Dotson(2019)}]{Dotson2019:pcs}
\bibinfo{author}{Dotson, C.}, \bibinfo{year}{2019}.
\newblock \bibinfo{title}{Practical Cloud Security: A Guide for Secure Design
  and Deployment}.
\newblock \bibinfo{edition}{1} ed., \bibinfo{publisher}{O'Reilly Media},
  \bibinfo{address}{Sebastopol, CA}.
\newblock \bibinfo{note}{ISBN 978-1-492-03751-4}.
\bibitem[{Dukhovni and Hardaker(2015)}]{rfc7671}
\bibinfo{author}{Dukhovni, V.}, \bibinfo{author}{Hardaker, W.},
  \bibinfo{year}{2015}.
\newblock \bibinfo{title}{The {DNS}-based authentication of named entities
  ({DANE}) protocol: Updates and operational guidance}.
\newblock \bibinfo{howpublished}{RFC 7671}.
\newblock \URLprefix \url{https://datatracker.ietf.org/doc/html/rfc7671}.
\bibitem[{Enck and Williams(2022)}]{enck2022top}
\bibinfo{author}{Enck, W.}, \bibinfo{author}{Williams, L.},
  \bibinfo{year}{2022}.
\newblock \bibinfo{title}{Top five challenges in software supply chain
  security: Observations from 30 industry and government organizations}.
\newblock \bibinfo{journal}{IEEE Security \& Privacy} \bibinfo{volume}{20},
  \bibinfo{pages}{96--100}.
\bibitem[{{ENISA}(2021)}]{enisa2021threat}
\bibinfo{author}{{ENISA}}, \bibinfo{year}{2021}.
\newblock \bibinfo{title}{{ENISA} threat landscape for supply chain attacks}.
\newblock \URLprefix
  \url{https://www.enisa.europa.eu/publications/threat-landscape-for-supply-chain-attacks/}.
\bibitem[{Fong et~al.(2016)Fong, Wheeler and Henninger}]{wheeler2016soar}
\bibinfo{author}{Fong, E.K.H.}, \bibinfo{author}{Wheeler, D.A.},
  \bibinfo{author}{Henninger, A.E.}, \bibinfo{year}{2016}.
\newblock \bibinfo{title}{{State-of-the-Art Resources (SOAR) for Software
  Vulnerability Detection, Test, and Evaluation 2016}}.
\newblock \bibinfo{type}{IDA Paper} \bibinfo{number}{P-8005}. Institute for
  Defense Analysis.
\bibitem[{Franceschi-Bicchierai(2021)}]{lorenzofranceschibicchierai2021}
\bibinfo{author}{Franceschi-Bicchierai, L.}, \bibinfo{year}{2021}.
\newblock \bibinfo{title}{The fortnite trial is exposing details about the
  biggest {iPhone} hack on record}.
\newblock \bibinfo{howpublished}{VICE}.
\newblock \URLprefix
  \url{https://www.vice.com/en/article/n7bbmz/the-fortnite-trial-is-exposing-details-about-the-biggest-iphone-hack-of-all-time}.
\bibitem[{Gallagher(2016)}]{gallagher2016}
\bibinfo{author}{Gallagher, S.}, \bibinfo{year}{2016}.
\newblock \bibinfo{title}{Rage-quit: Coder unpublished 17 lines of {JavaScript}
  and 'broke the internet'}.
\newblock \bibinfo{howpublished}{Ars Technica}.
\newblock \URLprefix
  \url{https://arstechnica.com/information-technology/2016/03/rage-quit-coder-unpublished-17-lines-of-javascript-and-broke-the-internet/}.
\bibitem[{Gallagher(2019)}]{gallagher2019appleinsider}
\bibinfo{author}{Gallagher, W.}, \bibinfo{year}{2019}.
\newblock \bibinfo{title}{Editorial: A year later, bloomberg silently stands by
  its 'big hack' icloud spy chip story}.
\newblock \bibinfo{howpublished}{AppleInsider}.
\newblock \URLprefix
  \url{https://appleinsider.com/articles/19/10/04/editorial-a-year-later-bloomberg-silently-stands-by-its-big-hack-icloud-spy-chip-story}.
\bibitem[{Gerste(2022)}]{gerste2022securing}
\bibinfo{author}{Gerste, P.}, \bibinfo{year}{2022}.
\newblock \bibinfo{title}{Securing developer tools: Package managers}.
\newblock \bibinfo{howpublished}{Sonar Blog}.
\newblock \URLprefix
  \url{https://blog.sonarsource.com/securing-developer-tools-package-managers}.
\bibitem[{{Git~SCM}(2022)}]{Git2022}
\bibinfo{author}{{Git~SCM}}, \bibinfo{year}{2022}.
\newblock \bibinfo{title}{Signing your work}.
\newblock \URLprefix
  \url{https://git-scm.com/book/en/v2/Git-Tools-Signing-Your-Work}.
\bibitem[{Goodin(2017)}]{goodin}
\bibinfo{author}{Goodin, D.}, \bibinfo{year}{2017}.
\newblock \bibinfo{title}{Devs unknowingly use ``malicious'' modules snuck into
  official python repository}.
\newblock \bibinfo{howpublished}{Ars Technica}.
\newblock \URLprefix
  \url{https://arstechnica.com/information-technology/2017/09/devs-unknowingly-use-malicious-modules-put-into-official-python-repository/}.
\bibitem[{Goodin(2019)}]{goodin2019supplychain}
\bibinfo{author}{Goodin, D.}, \bibinfo{year}{2019}.
\newblock \bibinfo{title}{The year-long rash of supply chain attacks against
  open source is getting worse}.
\newblock \bibinfo{howpublished}{Ars Technica}.
\newblock \URLprefix
  \url{https://arstechnica.com/information-technology/2019/08/the-year-long-rash-of-supply-chain-attacks-against-open-source-is-getting-worse/}.
\bibitem[{Goodin(2020)}]{goodin2020solarwinds}
\bibinfo{author}{Goodin, D.}, \bibinfo{year}{2020}.
\newblock \bibinfo{title}{$\sim$18,000 organizations downloaded backdoor
  planted by {Cozy Bear} hackers}.
\newblock \bibinfo{howpublished}{Ars Technica}.
\newblock \URLprefix
  \url{https://arstechnica.com/information-technology/2020/12/18000-organizations-downloaded-backdoor-planted-by-cozy-bear-hackers/}.
\bibitem[{Goodin(2021)}]{goodin2021maliciousNPM}
\bibinfo{author}{Goodin, D.}, \bibinfo{year}{2021}.
\newblock \bibinfo{title}{Malicious {NPM} packages are part of a malware
  ''barrage`` hitting repositories}.
\newblock \bibinfo{howpublished}{Ars Technica}.
\newblock \URLprefix
  \url{https://arstechnica.com/information-technology/2021/12/malicious-packages-sneaked-into-npm-repository-stole-discord-tokens/}.
\bibitem[{Goodin(2022)}]{goodin2022phishing}
\bibinfo{author}{Goodin, D.}, \bibinfo{year}{2022}.
\newblock \bibinfo{title}{Ongoing phishing campaign can hack you even when
  you’re protected with {MFA}}.
\newblock \bibinfo{howpublished}{Ars Technica}.
\newblock \URLprefix
  \url{https://arstechnica.com/information-technology/2022/07/microsoft-details-phishing-campaign-that-can-hijack-mfa-protected-accounts/}.
\bibitem[{Grassi et~al.(2017)Grassi, Fenton, Newton, Perlner, Regenscheid,
  Burr, Richer et~al.}]{grassi2017nist}
\bibinfo{author}{Grassi, P.A.}, \bibinfo{author}{Fenton, J.L.},
  \bibinfo{author}{Newton, E.M.}, \bibinfo{author}{Perlner, R.A.},
  \bibinfo{author}{Regenscheid, A.R.}, \bibinfo{author}{Burr, W.E.},
  \bibinfo{author}{Richer}, et~al., \bibinfo{year}{2017}.
\newblock \bibinfo{title}{{NIST} special publication 800-63b digital identity
  guidelines}.
\newblock \bibinfo{journal}{National Institute of Standards and Technology
  (NIST)} .
\bibitem[{Grimes(2020)}]{grimes2020hacking}
\bibinfo{author}{Grimes, R.A.}, \bibinfo{year}{2020}.
\newblock \bibinfo{title}{Hacking Multifactor Authentication}.
\newblock \bibinfo{publisher}{John Wiley \& Sons},
  \bibinfo{address}{Indianopolis, Indiana}.
\newblock \bibinfo{note}{ISBN 978-1-119-65080-5}.
\bibitem[{Hamid and Weber(2018)}]{hamid2018engineering}
\bibinfo{author}{Hamid, B.}, \bibinfo{author}{Weber, D.}, \bibinfo{year}{2018}.
\newblock \bibinfo{title}{Engineering secure systems: Models, patterns and
  empirical validation}.
\newblock \bibinfo{journal}{Computers \& Security} \bibinfo{volume}{77},
  \bibinfo{pages}{315--348}.
\newblock \DOIprefix\doi{https://doi.org/10.1016/j.cose.2018.03.016}.
\bibitem[{He et~al.(2015)He, Roe, Wood, Nachtigal and Helms}]{he2015model}
\bibinfo{author}{He, S.L.}, \bibinfo{author}{Roe, N.H.}, \bibinfo{author}{Wood,
  E.}, \bibinfo{author}{Nachtigal, N.M.}, \bibinfo{author}{Helms, J.},
  \bibinfo{year}{2015}.
\newblock \bibinfo{title}{Model of the Product Development Lifecycle}.
\newblock \bibinfo{type}{Technical Report}. Sandia National Lab.(SNL-NM),
  Albuquerque, NM (United States).
\newblock \URLprefix \url{https://www.osti.gov/biblio/1226426},
  \DOIprefix\doi{10.2172/1226426}.
\bibitem[{Hern(2014)}]{hern}
\bibinfo{author}{Hern, A.}, \bibinfo{year}{2014}.
\newblock \bibinfo{title}{Tor users advised to check their computers for
  malware}.
\newblock \bibinfo{howpublished}{The Guardian}.
\newblock \URLprefix
  \url{https://www.theguardian.com/technology/2014/oct/28/tor-users-advised-check-computers-malware}.
\bibitem[{Hernan et~al.(2006)Hernan, Lambert, Ostwald and
  Shostack}]{hernan2006threat}
\bibinfo{author}{Hernan, S.}, \bibinfo{author}{Lambert, S.},
  \bibinfo{author}{Ostwald, T.}, \bibinfo{author}{Shostack, A.},
  \bibinfo{year}{2006}.
\newblock \bibinfo{title}{Threat modeling - uncover security design flaws using
  the {STRIDE} approach}.
\newblock \bibinfo{journal}{MSDN Magazine-Louisville} ,
  \bibinfo{pages}{68--75}.
\bibitem[{Hoffman and Schlyter(2012)}]{rfc6698}
\bibinfo{author}{Hoffman, P.}, \bibinfo{author}{Schlyter, J.},
  \bibinfo{year}{2012}.
\newblock \bibinfo{title}{The {DNS}-based authentication of named entities
  ({DANE}) transport layer security ({TLS}) protocol: {TLSA}}.
\newblock \bibinfo{howpublished}{RFC 6698}.
\newblock \URLprefix \url{https://datatracker.ietf.org/doc/html/rfc6698}.
\bibitem[{Houser et~al.(2021)Houser, Hao, Li, Liu, Cotton and
  Wang}]{houser2021comprehensive}
\bibinfo{author}{Houser, R.}, \bibinfo{author}{Hao, S.}, \bibinfo{author}{Li,
  Z.}, \bibinfo{author}{Liu, D.}, \bibinfo{author}{Cotton, C.},
  \bibinfo{author}{Wang, H.}, \bibinfo{year}{2021}.
\newblock \bibinfo{title}{A comprehensive measurement-based investigation of
  {DNS} hijacking}, in: \bibinfo{booktitle}{40th International Symposium on
  Reliable Distributed Systems (SRDS)}, \bibinfo{organization}{IEEE}. pp.
  \bibinfo{pages}{210--221}.
\bibitem[{Humble and Farley(2010)}]{humble2010continuous}
\bibinfo{author}{Humble, J.}, \bibinfo{author}{Farley, D.},
  \bibinfo{year}{2010}.
\newblock \bibinfo{title}{Continuous Delivery: Reliable Software Releases
  through Build, Test, and Deployment Automation}.
\newblock \bibinfo{edition}{1} ed., \bibinfo{publisher}{Pearson Education}.
\bibitem[{Hunker and Probst(2011)}]{hunker2011insiders}
\bibinfo{author}{Hunker, J.}, \bibinfo{author}{Probst, C.W.},
  \bibinfo{year}{2011}.
\newblock \bibinfo{title}{Insiders and insider threats -- an overview of
  definitions and mitigation techniques}.
\newblock \bibinfo{journal}{Journal of Wireless Mobile Networks, Ubiquitous
  Computing and Dependable Applications} \bibinfo{volume}{2},
  \bibinfo{pages}{4--27}.
\bibitem[{{ISO 27002}(2013)}]{ISO2013ISO27002}
\bibinfo{author}{{ISO 27002}}, \bibinfo{year}{2013}.
\newblock \bibinfo{title}{ISO/IEC 27002:2013 Information technology -- Security
  techniques -- Code of practice for information security controls}.
  \bibinfo{edition}{2} ed.
\newblock \bibinfo{organization}{International Organization for
  Standardization}.
\bibitem[{Jelacic et~al.(2017)Jelacic, Rosic, Lendak, Stanojevic and
  Stoja}]{jelacic2017stride}
\bibinfo{author}{Jelacic, B.}, \bibinfo{author}{Rosic, D.},
  \bibinfo{author}{Lendak, I.}, \bibinfo{author}{Stanojevic, M.},
  \bibinfo{author}{Stoja, S.}, \bibinfo{year}{2017}.
\newblock \bibinfo{title}{{STRIDE} to a secure smart grid in a hybrid cloud},
  in: \bibinfo{booktitle}{Computer Security}. \bibinfo{publisher}{Springer,
  Cham}, pp. \bibinfo{pages}{77--90}.
\newblock \DOIprefix\doi{10.1007/978-3-319-72817-9_6}.
\bibitem[{Jenkins(2016)}]{jenkins2016}
\bibinfo{author}{Jenkins}, \bibinfo{year}{2016}.
\newblock \bibinfo{title}{Jenkins: Build great things at any scale}.
\newblock \bibinfo{howpublished}{Jenkins}.
\newblock \URLprefix \url{https://www.jenkins.io/}.
\bibitem[{Jover(2020)}]{jover2020sms}
\bibinfo{author}{Jover, R.P.}, \bibinfo{year}{2020}.
\newblock \bibinfo{title}{Security analysis of {SMS} as a second factor of
  authentication}.
\newblock \bibinfo{journal}{Communications of the ACM} \bibinfo{volume}{63},
  \bibinfo{pages}{46–--52}.
\newblock \DOIprefix\doi{10.1145/3424260}.
\bibitem[{Kan(2019)}]{michaelkan2019google}
\bibinfo{author}{Kan, M.}, \bibinfo{year}{2019}.
\newblock \bibinfo{title}{Google: Phishing attacks that can beat two-factor are
  on the rise}.
\newblock \bibinfo{howpublished}{PCMag}.
\newblock \URLprefix
  \url{https://www.pcmag.com/news/google-phishing-attacks-that-can-beat-two-factor-are-on-the-rise}.
\bibitem[{Karahasanovic et~al.(2017)Karahasanovic, Kleberger and
  Almgren}]{karahasanovic2017:adapting}
\bibinfo{author}{Karahasanovic, A.}, \bibinfo{author}{Kleberger, P.},
  \bibinfo{author}{Almgren, M.}, \bibinfo{year}{2017}.
\newblock \bibinfo{title}{Adapting threat modeling methods for the automotive
  industry}, in: \bibinfo{booktitle}{15th ESCAR Conference}, pp.
  \bibinfo{pages}{1--10}.
\bibitem[{Kent and Souppaya(2006)}]{kent2006:nist}
\bibinfo{author}{Kent, K.}, \bibinfo{author}{Souppaya, M.},
  \bibinfo{year}{2006}.
\newblock \bibinfo{title}{Guide to Computer Security Log Management}.
\newblock \bibinfo{type}{NIST SP} \bibinfo{number}{800-92}. National Institute
  of Standards and Technology.
\newblock \DOIprefix\doi{10.6028/NIST.SP.800-92}.
\bibitem[{Khan et~al.(2017)Khan, McLaughlin, Laverty and
  Sezer}]{Khan2017:stride}
\bibinfo{author}{Khan, R.}, \bibinfo{author}{McLaughlin, K.},
  \bibinfo{author}{Laverty, D.}, \bibinfo{author}{Sezer, S.},
  \bibinfo{year}{2017}.
\newblock \bibinfo{title}{{STRIDE}-based threat modeling for cyber-physical
  systems}, in: \bibinfo{booktitle}{2017 IEEE PES Innovative Smart Grid
  Technologies Conference Europe (ISGT-Europe)}, pp. \bibinfo{pages}{1--6}.
\newblock \DOIprefix\doi{10.1109/ISGTEurope.2017.8260283}.
\bibitem[{Kupsch et~al.(2017)Kupsch, Heymann, Miller and
  Basupalli}]{kupsch2017bad}
\bibinfo{author}{Kupsch, J.A.}, \bibinfo{author}{Heymann, E.},
  \bibinfo{author}{Miller, B.}, \bibinfo{author}{Basupalli, V.},
  \bibinfo{year}{2017}.
\newblock \bibinfo{title}{Bad and good news about using software assurance
  tools}.
\newblock \bibinfo{journal}{Software: Practice and Experience}
  \bibinfo{volume}{47}, \bibinfo{pages}{143--156}.
\bibitem[{Ladisa et~al.(2022)Ladisa, Plate, Martinez and
  Barais}]{ladisa2022taxonomy}
\bibinfo{author}{Ladisa, P.}, \bibinfo{author}{Plate, H.},
  \bibinfo{author}{Martinez, M.}, \bibinfo{author}{Barais, O.},
  \bibinfo{year}{2022}.
\newblock \bibinfo{title}{Taxonomy of attacks on open-source software supply
  chains}.
\newblock \bibinfo{journal}{arXiv preprint arXiv:2204.04008} \URLprefix
  \url{https://arxiv.org/abs/2204.04008}.
\bibitem[{Lamb and Zacchiroli(2022)}]{lamb2022reproducible}
\bibinfo{author}{Lamb, C.}, \bibinfo{author}{Zacchiroli, S.},
  \bibinfo{year}{2022}.
\newblock \bibinfo{title}{Reproducible builds: Increasing the integrity of
  software supply chains}.
\newblock \bibinfo{journal}{IEEE Software} \bibinfo{volume}{39},
  \bibinfo{pages}{62--70}.
\newblock \DOIprefix\doi{10.1109/MS.2021.3073045}.
\bibitem[{Le~Vie(2000)}]{le2000understanding}
\bibinfo{author}{Le~Vie, D.S.}, \bibinfo{year}{2000}.
\newblock \bibinfo{title}{Understanding data flow diagrams}, in:
  \bibinfo{booktitle}{Annual Conference-Society for Technical Communication},
  pp. \bibinfo{pages}{396--401}.
\bibitem[{Levy(2003)}]{levy2003:poisoning}
\bibinfo{author}{Levy, E.}, \bibinfo{year}{2003}.
\newblock \bibinfo{title}{Poisoning the software supply chain}.
\newblock \bibinfo{journal}{IEEE Security \& Privacy} \bibinfo{volume}{1},
  \bibinfo{pages}{70--73}.
\bibitem[{Ma and Schmittner(2016)}]{ma2016:threat}
\bibinfo{author}{Ma, Z.}, \bibinfo{author}{Schmittner, C.},
  \bibinfo{year}{2016}.
\newblock \bibinfo{title}{Threat modeling for automotive security analysis}.
\newblock \bibinfo{journal}{Advanced Science and Technology Letters}
  \bibinfo{volume}{139}, \bibinfo{pages}{333--339}.
\bibitem[{Maddox(2018)}]{maddox2018best}
\bibinfo{author}{Maddox, I.}, \bibinfo{year}{2018}.
\newblock \bibinfo{title}{12 best practices for user account, authentication
  and password management}.
\newblock \bibinfo{howpublished}{Google Cloud Platform}.
\newblock \URLprefix
  \url{https://cloud.google.com/blog/products/gcp/12-best-practices-for-user-account}.
\bibitem[{Marksteiner et~al.(2019)Marksteiner, Vallant and
  Nahrgang}]{Marksteiner2019:cybersec}
\bibinfo{author}{Marksteiner, S.}, \bibinfo{author}{Vallant, H.},
  \bibinfo{author}{Nahrgang, K.}, \bibinfo{year}{2019}.
\newblock \bibinfo{title}{Cyber security requirements engineering for
  low-voltage distribution smart grid architectures using threat modeling}.
\newblock \bibinfo{journal}{Journal of Information Security and Applications}
  \bibinfo{volume}{49}.
\newblock \DOIprefix\doi{10.1016/j.jisa.2019.102389}.
\bibitem[{Maunder(2017)}]{maunder2017psa}
\bibinfo{author}{Maunder, M.}, \bibinfo{year}{2017}.
\newblock \bibinfo{title}{{PSA: 4.8 Million Affected by Chrome Extension
  Attacks Targeting Site Owners}}.
\newblock \URLprefix
  \url{https://www.wordfence.com/blog/2017/08/chrome-browser-extension-attacks/}.
\bibitem[{McGraw(2012)}]{mcgraw2012software}
\bibinfo{author}{McGraw, G.}, \bibinfo{year}{2012}.
\newblock \bibinfo{title}{Software security}.
\newblock \bibinfo{journal}{Datenschutz und Datensicherheit--DuD}
  \bibinfo{volume}{36}, \bibinfo{pages}{662--665}.
\bibitem[{Miller(2013)}]{miller2013supply}
\bibinfo{author}{Miller, J.F.}, \bibinfo{year}{2013}.
\newblock \bibinfo{title}{Supply chain attack framework and attack patterns}.
\newblock \bibinfo{type}{Technical Report}. MITRE Corp.
  \bibinfo{address}{McLean, VA}.
\newblock \URLprefix \url{https://apps.dtic.mil/sti/citations/ADA610495}.
\bibitem[{Morris(2016)}]{morris2016infrastructure}
\bibinfo{author}{Morris, K.}, \bibinfo{year}{2016}.
\newblock \bibinfo{title}{Infrastructure as Code: Managing Servers in the
  Cloud}.
\newblock \bibinfo{publisher}{O'Reilly Media}.
\bibitem[{Myagmar et~al.(2005)Myagmar, Lee and Yurcik}]{myagmar2005threat}
\bibinfo{author}{Myagmar, S.}, \bibinfo{author}{Lee, A.J.},
  \bibinfo{author}{Yurcik, W.}, \bibinfo{year}{2005}.
\newblock \bibinfo{title}{{Threat Modeling as a Basis for Security
  Requirements}}, in: \bibinfo{booktitle}{Symposium on requirements engineering
  for information security (SREIS)}, \bibinfo{organization}{Citeseer}. pp.
  \bibinfo{pages}{1--8}.
\bibitem[{Möckel and Abdallah(2010)}]{mockel2010threat}
\bibinfo{author}{Möckel, C.}, \bibinfo{author}{Abdallah, A.E.},
  \bibinfo{year}{2010}.
\newblock \bibinfo{title}{Threat modeling approaches and tools for securing
  architectural designs of an e-banking application}, in:
  \bibinfo{booktitle}{2010 Sixth International Conference on Information
  Assurance and Security}, pp. \bibinfo{pages}{149--154}.
\newblock \DOIprefix\doi{10.1109/ISIAS.2010.5604049}.
\bibitem[{NIST(2019)}]{nist2019mfa}
\bibinfo{author}{NIST}, \bibinfo{year}{2019}.
\newblock \bibinfo{title}{{Back to basics: Multi-factor authentication (MFA)}}.
\newblock \bibinfo{howpublished}{NIST}.
\newblock \URLprefix
  \url{https://www.nist.gov/itl/applied-cybersecurity/tig/back-basics-multi-factor-authentication}.
\bibitem[{Nordstr{\"o}m and Dovrolis(2004)}]{nordstrom2004beware}
\bibinfo{author}{Nordstr{\"o}m, O.}, \bibinfo{author}{Dovrolis, C.},
  \bibinfo{year}{2004}.
\newblock \bibinfo{title}{Beware of {BGP} attacks}.
\newblock \bibinfo{journal}{ACM SIGCOMM Computer Communication Review}
  \bibinfo{volume}{34}, \bibinfo{pages}{1--8}.
\bibitem[{Ohm et~al.(2022)Ohm, Boes, Bungartz and Meier}]{ohm2022feasibility}
\bibinfo{author}{Ohm, M.}, \bibinfo{author}{Boes, F.},
  \bibinfo{author}{Bungartz, C.}, \bibinfo{author}{Meier, M.},
  \bibinfo{year}{2022}.
\newblock \bibinfo{title}{On the feasibility of supervised machine learning for
  the detection of malicious software packages}, in:
  \bibinfo{booktitle}{Proceedings of the 17th International Conference on
  Availability, Reliability and Security (ARES)}, pp. \bibinfo{pages}{1--10}.
\bibitem[{Ohm et~al.(2020)Ohm, Plate, Sykosch and Meier}]{ohm2020backstabber}
\bibinfo{author}{Ohm, M.}, \bibinfo{author}{Plate, H.},
  \bibinfo{author}{Sykosch, A.}, \bibinfo{author}{Meier, M.},
  \bibinfo{year}{2020}.
\newblock \bibinfo{title}{Backstabber’s knife collection: A review of open
  source software supply chain attacks}, in: \bibinfo{booktitle}{International
  Conference on Detection of Intrusions and Malware, and Vulnerability
  Assessment (DIMVA)}, \bibinfo{organization}{Springer}. pp.
  \bibinfo{pages}{23--43}.
\bibitem[{OWASP(2020a)}]{owaspSAMM2020}
\bibinfo{author}{OWASP}, \bibinfo{year}{2020}a.
\newblock \bibinfo{title}{{OWASP} software assurance maturity model, v2.0}.
\newblock \URLprefix \url{https://owaspsamm.org/}.
\bibitem[{OWASP(2020b)}]{owasp}
\bibinfo{author}{OWASP}, \bibinfo{year}{2020}b.
\newblock \bibinfo{title}{{Pinning Cheat Sheet}}.
\newblock \URLprefix
  \url{https://cheatsheetseries.owasp.org/cheatsheets/Pinning_Cheat_Sheet.html#pinning-cheat-sheet}.
\bibitem[{Owen(2018)}]{owen2018apple}
\bibinfo{author}{Owen, M.}, \bibinfo{year}{2018}.
\newblock \bibinfo{title}{Apple denies claim {China} slipped spy chips into its
  {iCloud} server hardware}.
\newblock \bibinfo{howpublished}{AppleInsider}.
\newblock \URLprefix
  \url{https://appleinsider.com/articles/18/10/04/apple-denies-claims-china-attacked-icloud-server-supply-chain-to-spy-on-us}.
\bibitem[{{Palo Alto Networks}(2020)}]{PaloAlto2020}
\bibinfo{author}{{Palo Alto Networks}}, \bibinfo{year}{2020}.
\newblock \bibinfo{title}{{Unit 42 Cloud Threat Report: Spring 2020}}.
\newblock \URLprefix
  \url{https://unit42.paloaltonetworks.com/cloud-threat-report-intro/}.
\bibitem[{Paule(2018)}]{paule2018securing}
\bibinfo{author}{Paule, C.}, \bibinfo{year}{2018}.
\newblock \bibinfo{title}{{Securing DevOps: detection of vulnerabilities in CD
  pipelines}}.
\newblock Master's thesis. Institute of Software Technology and University of
  Stuttgart.
\bibitem[{Peisert et~al.(2021)Peisert, Schneier, Okhravi, Massacci, Benzel,
  Landwehr, Mannan, Mirkovic, Prakash and Michael}]{peisert2021perspectives}
\bibinfo{author}{Peisert, S.}, \bibinfo{author}{Schneier, B.},
  \bibinfo{author}{Okhravi, H.}, \bibinfo{author}{Massacci, F.},
  \bibinfo{author}{Benzel, T.}, \bibinfo{author}{Landwehr, C.},
  \bibinfo{author}{Mannan, M.}, \bibinfo{author}{Mirkovic, J.},
  \bibinfo{author}{Prakash, A.}, \bibinfo{author}{Michael, J.B.},
  \bibinfo{year}{2021}.
\newblock \bibinfo{title}{Perspectives on the {SolarWinds} incident}.
\newblock \bibinfo{journal}{IEEE Security \& Privacy} \bibinfo{volume}{19},
  \bibinfo{pages}{7--13}.
\bibitem[{Pistoia et~al.(2007)Pistoia, Chandra, Fink and
  Yahav}]{pistoia2007survey}
\bibinfo{author}{Pistoia, M.}, \bibinfo{author}{Chandra, S.},
  \bibinfo{author}{Fink, S.J.}, \bibinfo{author}{Yahav, E.},
  \bibinfo{year}{2007}.
\newblock \bibinfo{title}{A survey of static analysis methods for identifying
  security vulnerabilities in software systems}.
\newblock \bibinfo{journal}{IBM Systems Journal} \bibinfo{volume}{46},
  \bibinfo{pages}{265--288}.
\newblock \DOIprefix\doi{10.1147/sj.462.0265}.
\bibitem[{Polkovnychenko and Menashe(2021)}]{polkovnychenko2021malicious}
\bibinfo{author}{Polkovnychenko, A.}, \bibinfo{author}{Menashe, S.},
  \bibinfo{year}{2021}.
\newblock \bibinfo{title}{Malicious npm packages are after your {Discord}
  tokens – 17 new packages disclosed}.
\newblock \bibinfo{howpublished}{JFrog}.
\newblock \URLprefix
  \url{https://jfrog.com/blog/malicious-npm-packages-are-after-your-discord-tokens-17-new-packages-disclosed/}.
\bibitem[{Purvine et~al.(2016)Purvine, Johnson and Lo}]{purvine2016graph}
\bibinfo{author}{Purvine, E.}, \bibinfo{author}{Johnson, J.R.},
  \bibinfo{author}{Lo, C.}, \bibinfo{year}{2016}.
\newblock \bibinfo{title}{A graph-based impact metric for mitigating lateral
  movement cyber attacks}, in: \bibinfo{booktitle}{Proceedings of the 2016 ACM
  Workshop on Automated Decision Making for Active Cyber Defense}, pp.
  \bibinfo{pages}{45--52}.
\bibitem[{Rahman et~al.(2019)Rahman, Parnin and Williams}]{rahman2019seven}
\bibinfo{author}{Rahman, A.}, \bibinfo{author}{Parnin, C.},
  \bibinfo{author}{Williams, L.}, \bibinfo{year}{2019}.
\newblock \bibinfo{title}{{The Seven Sins: Security Smells in Infrastructure as
  Code Scripts}}, in: \bibinfo{booktitle}{2019 IEEE/ACM 41st International
  Conference on Software Engineering (ICSE)}, pp. \bibinfo{pages}{164--175}.
\newblock \DOIprefix\doi{10.1109/ICSE.2019.00033}.
\bibitem[{Rahman et~al.(2021)Rahman, Rahman, Parnin and
  Williams}]{rahman2021security}
\bibinfo{author}{Rahman, A.}, \bibinfo{author}{Rahman, M.R.},
  \bibinfo{author}{Parnin, C.}, \bibinfo{author}{Williams, L.},
  \bibinfo{year}{2021}.
\newblock \bibinfo{title}{{Security Smells in {Ansible} and {Chef} Scripts: A
  Replication Study}}.
\newblock \bibinfo{journal}{ACM Transactions on Software Engineering and
  Methodology} \bibinfo{volume}{30}.
\bibitem[{Reed et~al.(2014)Reed, Miller and Popick}]{reed2014supply}
\bibinfo{author}{Reed, M.}, \bibinfo{author}{Miller, J.F.},
  \bibinfo{author}{Popick, P.}, \bibinfo{year}{2014}.
\newblock \bibinfo{title}{Supply chain attack patterns: Framework and catalog}.
\newblock \bibinfo{howpublished}{Office of the Deputy Assistant Secretary of
  Defense for Systems Engineering}.
\bibitem[{Rescorla(2018)}]{rfc8446}
\bibinfo{author}{Rescorla, E.}, \bibinfo{year}{2018}.
\newblock \bibinfo{title}{{The Transport Layer Security (TLS) Protocol Version
  1.3}}.
\newblock \bibinfo{howpublished}{RFC 8446 (Proposed Standard)}.
\newblock \URLprefix \url{https://tools.ietf.org/html/rfc8446},
  \DOIprefix\doi{10.17487/RFC8446}.
\bibitem[{Rimba et~al.(2015)Rimba, Zhu, Bass, Kuz and
  Reeves}]{rimba2015composing}
\bibinfo{author}{Rimba, P.}, \bibinfo{author}{Zhu, L.}, \bibinfo{author}{Bass,
  L.}, \bibinfo{author}{Kuz, I.}, \bibinfo{author}{Reeves, S.},
  \bibinfo{year}{2015}.
\newblock \bibinfo{title}{Composing patterns to construct secure systems}, in:
  \bibinfo{booktitle}{2015 11th European Dependable Computing Conference
  (EDCC)}, pp. \bibinfo{pages}{213--224}.
\newblock \DOIprefix\doi{10.1109/EDCC.2015.12}.
\bibitem[{Robertson and Riley(2018)}]{robertsonmichaelriley2018}
\bibinfo{author}{Robertson, J.}, \bibinfo{author}{Riley, M.},
  \bibinfo{year}{2018}.
\newblock \bibinfo{title}{{The Big Hack: How {China} Used a Tiny Chip to
  Infiltrate {U.S.} Companies}}.
\newblock \bibinfo{howpublished}{Bloomberg}.
\newblock \URLprefix
  \url{https://www.bloomberg.com/news/features/2018-10-04/the-big-hack-how-china-used-a-tiny-chip-to-infiltrate-america-s-top-companies}.
\bibitem[{Saltzer and Schroeder(1975)}]{saltzer1975:protection}
\bibinfo{author}{Saltzer, J.H.}, \bibinfo{author}{Schroeder, M.D.},
  \bibinfo{year}{1975}.
\newblock \bibinfo{title}{The protection of information in computer systems}.
\newblock \bibinfo{journal}{Proceedings of the IEEE} \bibinfo{volume}{63},
  \bibinfo{pages}{1278--1308}.
\bibitem[{Sanfilippo et~al.(2020)Sanfilippo, Abegaz, Payne and
  Salimi}]{sanfilippo2019}
\bibinfo{author}{Sanfilippo, J.}, \bibinfo{author}{Abegaz, T.},
  \bibinfo{author}{Payne, B.}, \bibinfo{author}{Salimi, A.},
  \bibinfo{year}{2020}.
\newblock \bibinfo{title}{{STRIDE}-based threat modeling for {MySQL}
  databases}, in: \bibinfo{editor}{Arai, K.}, \bibinfo{editor}{Bhatia, R.},
  \bibinfo{editor}{Kapoor, S.} (Eds.), \bibinfo{booktitle}{Proceedings of the
  Future Technologies Conference (FTC) 2019}, \bibinfo{publisher}{Springer
  International Publishing}. pp. \bibinfo{pages}{368--378}.
\bibitem[{Sattar et~al.(2021)Sattar, Vasoukolaei, Crysdale and
  Matrawy}]{sattar2021}
\bibinfo{author}{Sattar, D.}, \bibinfo{author}{Vasoukolaei, A.H.},
  \bibinfo{author}{Crysdale, P.}, \bibinfo{author}{Matrawy, A.},
  \bibinfo{year}{2021}.
\newblock \bibinfo{title}{A {STRIDE} threat model for {5G} core slicing}, in:
  \bibinfo{booktitle}{2021 IEEE 4th 5G World Forum (5GWF)}, pp.
  \bibinfo{pages}{247--252}.
\newblock \DOIprefix\doi{10.1109/5GWF52925.2021.00050}.
\bibitem[{Schmittner et~al.(2019)Schmittner, Tummeltshammer, Hofbauer, Shaaban,
  Meidlinger, Tauber, Bonitz, Hametner and
  Brandstetter}]{schmittner2019:threat}
\bibinfo{author}{Schmittner, C.}, \bibinfo{author}{Tummeltshammer, P.},
  \bibinfo{author}{Hofbauer, D.}, \bibinfo{author}{Shaaban, A.M.},
  \bibinfo{author}{Meidlinger, M.}, \bibinfo{author}{Tauber, M.},
  \bibinfo{author}{Bonitz, A.}, \bibinfo{author}{Hametner, R.},
  \bibinfo{author}{Brandstetter, M.}, \bibinfo{year}{2019}.
\newblock \bibinfo{title}{Threat modeling in the railway domain}, in:
  \bibinfo{booktitle}{International Conference on Reliability, Safety, and
  Security of Railway Systems}, \bibinfo{organization}{Springer, Cham}. pp.
  \bibinfo{pages}{261--271}.
\newblock \DOIprefix\doi{10.1007/978-3-030-18744-6_17}.
\bibitem[{Sharma(2022a)}]{sharma2022john}
\bibinfo{author}{Sharma, A.}, \bibinfo{year}{2022}a.
\newblock \bibinfo{title}{John {Deere} dependency confusion attempt flagged by
  {Sonatype}}.
\newblock \bibinfo{howpublished}{Sonatype}.
\newblock \URLprefix
  \url{https://blog.sonatype.com/john-deere-dependency-confusion-attempt-flagged-by-sonatype}.
\bibitem[{Sharma(2022b)}]{sharma2022pypi}
\bibinfo{author}{Sharma, A.}, \bibinfo{year}{2022}b.
\newblock \bibinfo{title}{{PyPI} flooded with 1,275 dependency confusion
  packages}.
\newblock \bibinfo{howpublished}{Sonatype}.
\newblock \URLprefix
  \url{https://blog.sonatype.com/pypi-flooded-with-over-1200-dependency-confusion-packages}.
\bibitem[{Sharma(2022c)}]{sharma2022python}
\bibinfo{author}{Sharma, A.}, \bibinfo{year}{2022}c.
\newblock \bibinfo{title}{python--dateutils -- a cryptominer in disguise
  targeting {Windows, Linux, macOS}}.
\newblock \bibinfo{howpublished}{Sonatype}.
\newblock \URLprefix
  \url{https://blog.sonatype.com/python-dateutils-a-moner-cryptominer-in-disguise-for-windows-linux-macos}.
\bibitem[{Sharma(2022d)}]{sharma2022string}
\bibinfo{author}{Sharma, A.}, \bibinfo{year}{2022}d.
\newblock \bibinfo{title}{{StringJS} typosquat deploys discord infostealer
  obfuscated five times}.
\newblock \bibinfo{howpublished}{Sonatype}.
\newblock \URLprefix
  \url{https://blog.sonatype.com/stringjs-typosquat-caught-with-discord-info-stealer}.
\bibitem[{Sharwood(2020)}]{sharwood2020}
\bibinfo{author}{Sharwood, S.}, \bibinfo{year}{2020}.
\newblock \bibinfo{title}{{OpenStack haven OpenDev yanks Gerrit code review
  tool after admin account compromised for two weeks}}.
\newblock \bibinfo{howpublished}{The Register}.
\newblock \URLprefix
  \url{https://www.theregister.com/2020/10/21/opendev_gerrit_attack/}.
\bibitem[{Shaw(2017)}]{shaw}
\bibinfo{author}{Shaw, R.A.}, \bibinfo{year}{2017}.
\newblock \bibinfo{title}{Software supply chain attacks}.
\newblock \URLprefix
  \url{https://csrc.nist.gov/CSRC/media/Projects/Supply-Chain-Risk-Management/documents/ssca/2017-winter/NCSC_Placemat.pdf}.
\bibitem[{Shevchenko et~al.(2018)Shevchenko, Chick, O'Riordan, Scanlon and
  Woody}]{shevchenko2018threat}
\bibinfo{author}{Shevchenko, N.}, \bibinfo{author}{Chick, T.A.},
  \bibinfo{author}{O'Riordan, P.}, \bibinfo{author}{Scanlon, T.P.},
  \bibinfo{author}{Woody, C.}, \bibinfo{year}{2018}.
\newblock \bibinfo{title}{Threat modeling: a summary of available methods}.
\newblock \bibinfo{type}{Technical Report}. Carnegie Mellon University Software
  Engineering Institute Pittsburgh United States.
\bibitem[{Shirey(2007)}]{rfc4949}
\bibinfo{author}{Shirey, R.W.}, \bibinfo{year}{2007}.
\newblock \bibinfo{title}{{Internet Security Glossary, Version 2}}.
\newblock \bibinfo{howpublished}{RFC 4949}.
\newblock \URLprefix \url{https://rfc-editor.org/rfc/rfc4949.txt},
  \DOIprefix\doi{10.17487/RFC4949}.
\bibitem[{Shostack(2014)}]{shostack2014threat}
\bibinfo{author}{Shostack, A.}, \bibinfo{year}{2014}.
\newblock \bibinfo{title}{Threat modeling: Designing for security}.
\newblock \bibinfo{publisher}{John Wiley \& Sons},
  \bibinfo{address}{Indianopolis, Indiana}.
\newblock \bibinfo{note}{ISBN 978-1-118-80999-0}.
\bibitem[{Simpson(2010)}]{simpson2010software}
\bibinfo{author}{Simpson, S.}, \bibinfo{year}{2010}.
\newblock \bibinfo{title}{Software integrity controls--an assurance--based
  approach to minimizing risks in the software supply chain}.
\newblock \bibinfo{type}{Technical Report}. SAFECode.
\bibitem[{Skrimstad(2018)}]{skrimstad2018improving}
\bibinfo{author}{Skrimstad, Y.}, \bibinfo{year}{2018}.
\newblock \bibinfo{title}{Improving Trust in Software through Diverse
  Double--Compiling and Reproducible Builds}.
\newblock Master's thesis. University of Oslo.
\bibitem[{SPIFFE(2022)}]{spiffe2022}
\bibinfo{author}{SPIFFE}, \bibinfo{year}{2022}.
\newblock \bibinfo{title}{{SPIFFE}: Secure production identity framework for
  everyone}.
\newblock \bibinfo{howpublished}{SPIFFE}.
\newblock \URLprefix \url{https://spiffe.io/}.
\bibitem[{{The Linux Foundation}(2020)}]{thelinuxfoundation2020}
\bibinfo{author}{{The Linux Foundation}}, \bibinfo{year}{2020}.
\newblock \bibinfo{title}{Open source software supply chain security}.
\newblock \URLprefix
  \url{https://www.linuxfoundation.org/resources/publications/open-source-software-supply-chain-security}.
\bibitem[{Theis et~al.(2019)Theis, Trzeciak, Costa, Moore, Miller, Cassidy and
  Claycomb}]{TheisCommonSense2019}
\bibinfo{author}{Theis, M.}, \bibinfo{author}{Trzeciak, R.},
  \bibinfo{author}{Costa, D.}, \bibinfo{author}{Moore, A.},
  \bibinfo{author}{Miller, S.}, \bibinfo{author}{Cassidy, T.},
  \bibinfo{author}{Claycomb, W.}, \bibinfo{year}{2019}.
\newblock \bibinfo{title}{Common Sense Guide to Mitigating Insider Threats}.
\newblock \bibinfo{type}{Technical Report}
  \bibinfo{number}{CMU/SEI-2018-TR-010}. Software Engineering Institute,
  Carnegie Mellon University. \bibinfo{address}{Pittsburgh, PA}.
\newblock \URLprefix
  \url{http://resources.sei.cmu.edu/library/asset-view.cfm?AssetID=540644}.
\bibitem[{Thompson(1984)}]{thompson1984reflections}
\bibinfo{author}{Thompson, K.}, \bibinfo{year}{1984}.
\newblock \bibinfo{title}{Reflections on trusting trust}.
\newblock \bibinfo{journal}{Communications of the ACM} \bibinfo{volume}{27},
  \bibinfo{pages}{761--763}.
\bibitem[{Torres-Arias et~al.(2019)Torres-Arias, Afzali, Kuppusamy, Curtmola
  and Cappos}]{torres2019toto}
\bibinfo{author}{Torres-Arias, S.}, \bibinfo{author}{Afzali, H.},
  \bibinfo{author}{Kuppusamy, T.K.}, \bibinfo{author}{Curtmola, R.},
  \bibinfo{author}{Cappos, J.}, \bibinfo{year}{2019}.
\newblock \bibinfo{title}{in-toto: Providing farm-to-table guarantees for bits
  and bytes}, in: \bibinfo{booktitle}{28th {USENIX} Security Symposium
  ({USENIX} Security 19)}, \bibinfo{publisher}{{USENIX} Association},
  \bibinfo{address}{Santa Clara, CA}. pp. \bibinfo{pages}{1393--1410}.
\newblock \URLprefix
  \url{https://www.usenix.org/conference/usenixsecurity19/presentation/torres-arias}.
\bibitem[{Tuma and Scandariato(2018)}]{tuma2018two}
\bibinfo{author}{Tuma, K.}, \bibinfo{author}{Scandariato, R.},
  \bibinfo{year}{2018}.
\newblock \bibinfo{title}{Two architectural threat analysis techniques
  compared}, in: \bibinfo{booktitle}{European Conference on Software
  Architecture}, \bibinfo{publisher}{Springer, Cham}. pp.
  \bibinfo{pages}{347--363}.
\newblock \DOIprefix\doi{10.1007/978-3-030-00761-4_23}.
\bibitem[{Vaidya et~al.(2019)Vaidya, Torres-Arias, Curtmola and
  Cappos}]{vaidya2019commit}
\bibinfo{author}{Vaidya, S.}, \bibinfo{author}{Torres-Arias, S.},
  \bibinfo{author}{Curtmola, R.}, \bibinfo{author}{Cappos, J.},
  \bibinfo{year}{2019}.
\newblock \bibinfo{title}{Commit signatures for centralized version control
  systems}, in: \bibinfo{booktitle}{IFIP International Conference on ICT
  Systems Security and Privacy Protection}, \bibinfo{organization}{Springer}.
  pp. \bibinfo{pages}{359--373}.
\bibitem[{Verissimo et~al.(2003)Verissimo, Neves and Correia}]{Verissimo2003}
\bibinfo{author}{Verissimo, P.}, \bibinfo{author}{Neves, N.F.},
  \bibinfo{author}{Correia, M.P.}, \bibinfo{year}{2003}.
\newblock \bibinfo{title}{Intrusion-tolerant architectures: Concepts and
  design}, in: \bibinfo{booktitle}{Architecting Dependable Systems}. volume
  \bibinfo{volume}{2677} of \textit{\bibinfo{series}{LNCS}}, pp.
  \bibinfo{pages}{3--36}.
\bibitem[{Wang et~al.(2013)Wang, Al~Sabbagh and Kowalski}]{wang2013socio}
\bibinfo{author}{Wang, X.}, \bibinfo{author}{Al~Sabbagh, B.},
  \bibinfo{author}{Kowalski, S.}, \bibinfo{year}{2013}.
\newblock \bibinfo{title}{A socio-technical framework for threat modeling a
  software supply chain}, in: \bibinfo{booktitle}{The 2013 Dewald Roode
  Workshop on Information Systems Security Research (Paper~17)},
  \bibinfo{organization}{International Federation for Information Processing}.
\bibitem[{Warren(2017)}]{tomwarren2017}
\bibinfo{author}{Warren, T.}, \bibinfo{year}{2017}.
\newblock \bibinfo{title}{Hackers hid malware in {CCleaner} software}.
\newblock \bibinfo{howpublished}{The Verge}.
\newblock \URLprefix
  \url{https://www.theverge.com/2017/9/18/16325202/ccleaner-hack-malware-security}.
\bibitem[{Webmin(2019)}]{webmin2019}
\bibinfo{author}{Webmin}, \bibinfo{year}{2019}.
\newblock \bibinfo{title}{Webmin 1.890 exploit - what happened?}
\newblock \bibinfo{howpublished}{Webmin}.
\newblock \URLprefix \url{https://www.webmin.com/exploit.html}.
\bibitem[{Weinert(2019)}]{weinert2019password}
\bibinfo{author}{Weinert, A.}, \bibinfo{year}{2019}.
\newblock \bibinfo{title}{Your pa\$\$word doesn't matter}.
\newblock \bibinfo{howpublished}{Microsoft}.
\newblock \URLprefix
  \url{https://techcommunity.microsoft.com/t5/azure-active-directory-identity/your-pa-word-doesn-t-matter/ba-p/731984}.
\bibitem[{Whalen(2001)}]{whalen2001:arp-spoofing}
\bibinfo{author}{Whalen, S.}, \bibinfo{year}{2001}.
\newblock \bibinfo{title}{An introduction to {ARP} spoofing}.
\newblock \bibinfo{howpublished}{Node99}.
\newblock \URLprefix
  \url{https://dl.packetstormsecurity.net/papers/protocols/intro_to_arp_spoofing.pdf}.
\bibitem[{Wheeler(2009)}]{wheeler2009fully}
\bibinfo{author}{Wheeler, D.A.}, \bibinfo{year}{2009}.
\newblock \bibinfo{title}{{Fully Countering Trusting Trust through Diverse
  Double-Compiling (DDC)}}.
\newblock Ph.D. thesis. George Mason University.
\bibitem[{Wheeler et~al.(2018)Wheeler, Reddy and Fong}]{wheeler2018:swa}
\bibinfo{author}{Wheeler, D.A.}, \bibinfo{author}{Reddy, D.J.},
  \bibinfo{author}{Fong, E.K.}, \bibinfo{year}{2018}.
\newblock \bibinfo{title}{{Securely Using Software Assurance ({SwA}) Tools in
  the Software Development Environment}}.
\newblock \bibinfo{type}{IDA Document} \bibinfo{number}{P-9166}. Institute for
  Defense Analysis.
\bibitem[{Wilhjelm and Younis(2020)}]{Wilhjelm2020}
\bibinfo{author}{Wilhjelm, C.}, \bibinfo{author}{Younis, A.A.},
  \bibinfo{year}{2020}.
\newblock \bibinfo{title}{A threat analysis methodology for security
  requirements elicitation in machine learning based systems}, in:
  \bibinfo{booktitle}{2020 IEEE 20th International Conference on Software
  Quality, Reliability and Security Companion (QRS-C)}, pp.
  \bibinfo{pages}{426--433}.
\newblock \DOIprefix\doi{10.1109/QRS-C51114.2020.00078}.
\bibitem[{Worrall(2015)}]{Juniper2015}
\bibinfo{author}{Worrall, B.}, \bibinfo{year}{2015}.
\newblock \bibinfo{title}{Important announcement about {ScreenOS}}.
\newblock \URLprefix
  \url{https://forums.juniper.net/t5/Security-Incident-Response/Important-Announcement-about-ScreenOS/ba-p/285554}.
\bibitem[{Xiao(2015)}]{claudxiao2015}
\bibinfo{author}{Xiao, C.}, \bibinfo{year}{2015}.
\newblock \bibinfo{title}{Novel malware {XcodeGhost} modifies {Xcode}, infects
  {Apple iOS} apps and hits app store}.
\newblock \bibinfo{howpublished}{Palo Alto Networks}.
\newblock \URLprefix
  \url{https://unit42.paloaltonetworks.com/novel-malware-xcodeghost-modifies-xcode-infects-apple-ios-apps-and-hits-app-store/}.

\end{thebibliography}
